\def\norm[#1]{\lVert{#1} \rVert}
\def\speedOfLight{\rm c}
\def\distance{R_{\indexUser,\indexLED}}
\def\Ar{A_{\rm R}}
\def\FOV{\theta_{\rm FOV}}
\def\mode{\gamma}
\def\angleLOSandTX{\phi_{\indexUser,\indexLED}}
\def\angleLOSandRX{\theta_{\indexUser,\indexLED}}
\def\rect[#1]{{\Pi}\left(#1\right)}
\def\dirac[#1]{{\delta}\left(#1\right)}

\def\realNumbers{\mathbb{R}}
\def\location[#1][#2]{{\rm \bf r}_{#1}^{#2}}
\def\orientation[#1][#2]{{\rm \bf q}_{#1}^{#2}}
\def\channelAtReflection[#1][#2]{h^{(#1)}(#2)}
\def\channell[#1]{h(#1)}
\def\source{\mathcal{S}_{\indexLED}}
\def\receiver{\mathcal{R}_{\indexUser}}
\def\degreeOfReflection{d}
\def\Cmatrix[#1]{{\bf C}(#1)}
\def\Dmatrix{{\bf D}}
\def\numberOfReflectors{L}
\def\frequency{f}
\def\pVector{{\bf p}}
\def\initialVector{{\bf e}}
\def\propagationDelay{\tau}
\def\area{A}
\def\deltaArea{{\rm d}\area}
\def\reflectionCoefficient[#1]{{\rho}_{#1}}
\def\eigenMatrix{{\bf Q}}
\def\eigenDiagonal{{\bf \Lambda}}

\def\NumberOfUsers{K}
\def\NumberOfLEDs{N}
\def\indexUser{k}
\def\indexLED{n}
\def\indexPD{m}
\def\NumberOfPDs{M}
\def\LEDGroup[#1]{l(#1)}
\def\LEDGroupNumber[#1]{l_{#1}}
\def\LEDConnectFlag[#1][#2]{\alpha_{#1#2}}
\def\channel[#1][#2]{h_{#1#2}}
\def\channelVector[#1][#2]{\textbf{h}_{#1#2}}
\def\power[#1]{p_{#1}}
\def\powerVector{\textbf{p}}
\def\ReceivedPower[#1][#2][#3]{H_{#3,#1(#2)}}
\def\signalVector[#1][#2]{\textbf{v}_{#2#1}}

\def\unassignedLEDs{\mathcal{U}}
\def\assignedto{\mathcal{T}_{\indexUser}}
\def\PDresp{r~\!\!}

\documentclass[journal,12pt,onecolumn,draftclsnofoot,]{IEEEtran}

\usepackage{xcolor}
\usepackage{acronym}
\usepackage{cite}
\usepackage{amsmath,amssymb}
\usepackage{graphicx}
\usepackage{subfigure}
\usepackage{caption}
\usepackage{algorithm,algcompatible}
\usepackage{gensymb}
\usepackage[keeplastbox]{flushend}

\usepackage{amsthm}%
\usepackage{etoolbox}%
\newtheorem{remark}{Remark}%
\AtEndEnvironment{remark}{\null\hfill\qedsymbol}%

\algdef{SE}[SUBALG]{Indent}{EndIndent}{}{\algorithmicend\ }%
\algtext*{Indent}
\algtext*{EndIndent}
\let\svthefootnote\thefootnote
\theoremstyle{remark}

\newtheorem{theorem}{Theorem}
\newtheorem{corollary}{Corollary}
\newtheorem{case}{Case}

\begin{document}

\title{\huge Multi-Element VLC Networks: LED Assignment, Power Control, and Optimum Combining}

\author{\IEEEauthorblockN{Yusuf Said Ero\u{g}lu, {\.I}smail G\"{u}ven\c{c}, \textit{Senior Member, IEEE,} Alphan \c{S}ahin, Yavuz~Yap{\i}c{\i}, Nezih~Pala and Murat Y\"{u}ksel}, \textit{Senior Member, IEEE}}

\maketitle
\thispagestyle{empty}

\acrodef{VLC}{Visible light communication}
\acrodef{CIR}{channel impulse response}
\acrodef{SIR}{signal-to-interference ratio}
\acrodef{SNR}{signal-to-noise ratio}
\acrodef{SINR}{signal-to-interference-plus-noise ratio}
\acrodef{PSD}{power spectral densitie}
\acrodef{MIMO}{multiple-input multiple-output}
\acrodef{MMSE}{minimum mean square error}
\acrodef{RF}{radio frequency}
\acrodef{VAP}{visible light access point} 
\acrodef{PA}{power amplifier}
\acrodef{BER}{bit error rate}
\acrodef{LS}{least squares}
\acrodef{LTE}{Long Term Evolution}
\acrodef{AWGN}{additive white Gaussian noise}
\acrodef{FIR}{finite impulse response}
\acrodef{LED}{light emitting diode}
\acrodef{FOV}{field-of-view}
\acrodef{PD}{photo detector}
\acrodef{RSS}{received signal strength}
\acrodef{LOS}{line-of-sight}
\acrodef{NLOS}{non-line-of-sight}
\acrodef{MSE}[MSE]{mean square error}
\acrodef{RMSE}[RMSE]{root mean square error}
\acrodef{ML}[ML]{Maximum likelihood}
\acrodef{NLS}[NLS]{nonlinear least squares}
\acrodef{FIM}{Fisher Information Matrix}
\acrodef{GPS}{Global Positioning System}
\acrodef{UWB}{ultra wideband}
\acrodef{WLAN}{wireless local area network}

\begin{abstract}
Visible light communications (VLC) is a promising technology to address the spectrum crunch problem in radio frequency (RF) networks. A major advantage of VLC networks is that they can use the existing lighting infrastructure in indoor environments, which may have large number of LEDs for illumination. While LEDs used for lighting typically have limited bandwidth, presence of many LEDs can be exploited for indoor VLC networks, to serve each user by multiple LEDs for improving link quality and throughput. In this paper, LEDs are grouped and assigned to the users based on received signal strength from each LED, for which different solutions are proposed to achieve maximum throughput, proportional fairness and quality of service (QoS). Additionally, power optimization of LEDs for a given assignment is investigated, and Jacobian and Hessian matrices of the corresponding optimization problem are derived. Moreover, for multi-element receivers with LED grouping at the transmitter, an improved optimal combining method is proposed. This method suppresses interference caused by simultaneous data transfer of LEDs and improves the overall signal-to-interference-plus-noise-ratio (SINR) by 2~dB to 5~dB. Lastly, an efficient calculation of channel response is presented to simulate multipath VLC channel with low computational complexity.
\end{abstract}


\section{Introduction}
\let\thefootnote\relax\footnote{ Yusuf Said Ero\u{g}lu, {\.I}smail G\"{u}ven\c{c} and Yavuz~Yap{\i}c{\i} are with the Department of Electrical and Computer Engineering at the North Carolina State University, Raleigh, NC. \{email: \{yeroglu, iguvenc, yyapici\}@ncsu.edu\}}
\let\thefootnote\relax\footnote{Alphan \c{S}ahin is with InterDigital Communications Inc., Huntington Quadrangle, Melville, NY. \{email: alphan.sahin@interdigital.com\}}
\let\thefootnote\relax\footnote{Nezih Pala is with the Department of Electrical and Computer Engineering at Florida International University, Miami, FL. \{email: npala@fiu.edu\}}
\let\thefootnote\relax\footnote{Murat Y\"{u}ksel is with the Department of Electrical and Computer Engineering at University of Central Florida, Orlando, FL. \{murat.yuksel@ucf.edu\}}
\let\thefootnote\relax\footnote{This work is supported in part by NSF CNS awards 1422354 and 1422062, ARO DURIP award W911NF-14-1-0531, and NASA NV Space Consortium.} \!\!\!\!
\addtocounter{footnote}{-5}\let\thefootnote\svthefootnote
Light emmiting diodes (LEDs) have become increasingly popular within the last decade as light sources due to their decreasing cost, low energy use and compatibility with different applications such as dimming. One of the most important applications enabled by the proliferation of LEDs is visible light communications (VLC) which transmit data in the visible light spectrum. Light intensity of LEDs can be modulated with a high frequency, which allows data transmission that is not perceivable to human eye \cite{rajagopal2012ieee}. VLC takes advantage of the light that is used for illumination as a communication channel and does not require additional signal for data transmission, which provides an opportunity for energy efficiency. Moreover, since the light is directional and does not penetrate walls, it allows high spatial reuse and provides inherent physical security \cite{6883837}. VLC is also considered to be a complementary technology to radio frequency (RF) technologies such as LTE and Wi-Fi \cite{6162563}. 

A main disadvantage of VLC is that the achievable data rates with commercial LEDs are not high due to their lower bandwidth. Phosphorescent white LEDs are reported to provide up to 20 MHz bandwidth, which enables up to 1~Gb/s data rates for a single user \cite{6249713}. In order to support multiple users and mobile applications that require significantly higher data rates and longer coverage range, advanced transceiver techniques are required. For this purpose, in this paper, we study three different approaches that can improve throughput for multi-element VLC networks.

Firstly, we study the assignment of multiple LEDs to users in a multi-LED transmitter VLC network. The number of LEDs is assumed to be much larger than the number of users, and multiple highly directional LEDs are assigned to each user. Using this scheme, simultaneous transmission to multiple users and hence higher signal-to-interference-plus-noise-ratio (SINR) are aimed for. A possible method is assigning LEDs to users based on their locations. However, in VLC networks, location information itself might not be sufficient. VLC connectivity is highly dependent on direct \ac{LOS} signals and the received signal strength (RSS) may decrease dramatically in case of an obstacle between the receiver and the user. Also, angle of arrival of a signal and a receiver's orientation can significantly affect the signal strength. Therefore, we directly use the RSS information at receiver to assign LEDs to users to provide a robust assignment scheme against obstructions. LED assignment is studied for two different scenarios, preallocated QoS rates and opportunistic sum rate maximization. We examine the tractability of finding an optimal assignment algorithm in both cases, and propose heuristic algorithms that find suboptimal solutions with low computational complexity.

Secondly, we optimize the transmit powers of the LEDs to improve the sum rate and fairness for a given LED to user assignment. We formulate the corresponding optimization problem, present Lagrangian dual function, and derive Jacobian and Hessian matrices.

Lastly, we utilize multi-element receiver diversity using optimal combining (OC), and propose a novel correlation matrix calculation method to capture the correlation between interference signals more accurately. The OC proposed in~\cite{1146095} is shown to provide higher SINR than its counterparts. It uses a correlation matrix of interferences received from different \acp{PD} to suppress interference. Assigning many LEDs having different locations and directions to the same user as in our proposed scheme creates a more diverse channel. If the user knows which LEDs are assigned together, this information can help to suppress interference more successfully. Simulation results show that the proposed OC calculation method in this paper improves the SINR by 2~dB to 5~dB over the OC proposed in \cite{1146095} at no additional cost or computational complexity.

The rest of this paper is organized as follows. Section~II presents a brief literature review, while Section~III introduces the system model and establishes a fast method of generating multipath realizations for VLC simulations. Section~IV introduces the LED assignment problem, proposes heuristic solution algorithms, and optimizes the transmit power of the LEDs, Section~V discusses calculation of optimal combining with LED grouping, Section~VI presents simulation results, and finally Section~VII concludes the paper.

\section{Literature Review}
In VLC networks, in order to provide ubiquitous illumination/wireless coverage, improve link quality, and provide higher throughput, a large number of LEDs with directional propagation characteristics can be deployed in indoor environments. Therefore, there might be significantly larger number of LEDs than the number of users in the network, and it is possible to serve each user with multiple directional LEDs over the same bandwidth \cite{Sahin:15, Nakagawa, 5342325, 7390984, eroglu_GlobecomWS, eroglu_VLCS,yuksel_WN_2009}. In RF communications, on the other hand, a single transmitter may serve many users at the same time simultaneously, and users may connect to the transmitter with the highest received power~\cite{4146798, 1400034}. Coordinated multipoint transmission (CoMP) technique offers serving a user with more than one base station~\cite{6146494, 6472200}, which is used in LTE-Advanced for interference management. However, typically, the number of users connected to a base station is significantly higher than the number of antennas deployed at the radio. The assignment techniques developed for RF communications are hence not directly applicable to VLC networks. The problem of assigning LEDs to users has similarities with the subchannel allocation in multiuser OFDM systems \cite{rhee2000increase, shen2005adaptive}. However, while assigning extra subchannel to a user affects its bandwidth, assigning an extra LED to a VLC user affects its SINR, and hence the impact on the system performance will be different.

In \cite{6825144, 7147818}, joint transmission of information from multiple VLC nodes is studied, in order to mitigate cell edge interference. An LED allocation scheme is presented in \cite{Sewaiwar}, which proposes an LED array as VLC transmitter, assigning LEDs with respect to the number and location of the users. However, it only provides a few fixed assignment options and cannot be generalized to transmitter deployments with different geometries. The studies on resource allocation for VLC networks \cite{bykhovsky2014multiple, zhang2012optimal}, reported in the literature focus on power allocation of LEDs and do not address how to assign multiple LEDs to a single user.

In \cite{eroglu_VLCS}, a multi element transmitter structure is studied for VLC, where multiple LEDs directed to different angles are placed on a light bulb. It is shown that multi-element transmitters serve more users simultaneously and provides more spatial reuse compared to transmitters with less directional LEDs. In \cite{SDMA}, a spatial division multiple access (SDMA) scheme is described to serve multiple users simultaneously by allocating them different LEDs, where the LED allocation is decided based on the location of the user. However, as mentioned in the introduction, location information itself might not always be sufficient to assign LEDs to users, since the RSS can dramatically decrease in case of an obstruction or a change in the receiver orientation.

In addition to the use of multi-element LED assignments to users, using multiple PDs at the receiver side is shown to improve the SINR performance~\cite{haas_multireceiver}. {The studies on imaging angle diversity receiver for infrared communications have been a baseline for VLC related extensions in this area~\cite{735884,848557,891218}. In \cite{891218}, the performance of angular diversity receivers with multibeam infrared transmitters is investigated and it is concluded that diversity reduces the ambient noise dramatically when used with maximum ratio combining (MRC).} Alternatively, the minimum-mean-squared-error interference-rejection-combining (MMSE-IRC) has been popular in RF MIMO communications for improving SINR under interference, and included in LTE standard with release 11~\cite{sagae2013}. The MMSE-IRC is able to use the multiple receiver antennas to create a null in the direction of the interference (where antenna gain significantly drops), and therefore suppress the interference. 

The OC is another related technique \cite{1146095} that considers the correlation between interference signals received from different PDs, and it suppresses the interference by using optimum combining weights. In~\cite{haas_multireceiver}, it is shown that the OC provides the highest SINR for multi-PD VLC and it is followed by the MRC, which weighs the PDs proportionally with their individual SINR measurements. In \cite{eroglu_VLCS}, it is shown that the OC provides even higher gain in comparison to MRC when multi-LED transmitters, as also considered in this paper, are used. However, these studies do not consider grouping of LEDs to serve users. When LEDs are grouped, interfering LEDs will also be grouped to serve other users, which will further increase the correlation between interference signals. In this case, interference correlation matrix modeling of OC also needs to change to. 

\section{System Model}

This section presents the network model, the VLC channel model, and the proposed method for efficient calculation of channel response.

\begin{figure}[tb]
	\centering
	\includegraphics[width = 3.5 in]{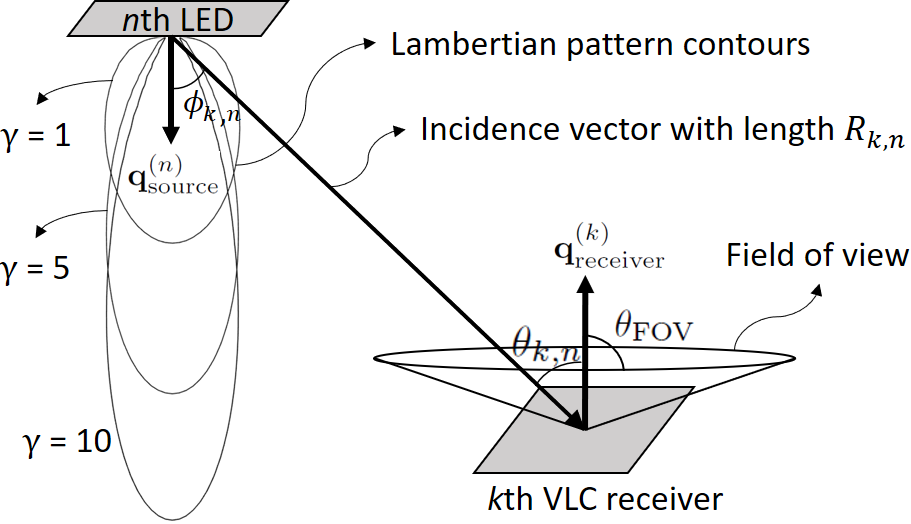}
	\caption{\small The LOS channel model with an LED and a receiver, and Lambertian pattern contours for different $\mode$. All the locations on the contours have identical RSS values. }
	\label{LambertianFigure}
\end{figure}

\subsection{VLC Channel Model}
We consider a multipath propagation environment in this paper based on the well-established models in the literature for \ac{LOS} and \ac{NLOS} scenarios. In order to characterize locations, orientations, and directionality of the LEDs, without loss of generality, the $\indexLED$th LED $\source$ can be defined with three parameters as $\source~=~\{\location[{\rm tx}][(\indexLED)],\orientation[{\rm tx}][(\indexLED)],\mode \},$
where $\location[{\rm tx}][(\indexLED)]\in\realNumbers^{3\times1}$ is the location of the $\indexLED$th LED,  $\orientation[{\rm tx}][(\indexLED)]\in\realNumbers^{3\times1}$ is the orientation of $\indexLED$th LED, and $\mode$ is the parameter that specifies the directionality of the light source based on Lambertian pattern as illustrated in Fig.~\ref{LambertianFigure}. Similarly, the $\indexUser$th receiver is modeled as $\receiver~=~\{\location[{\rm rx}][(\indexUser)],\orientation[{\rm rx}][(\indexUser)],\Ar, \FOV \}~,$ where $\location[{\rm rx}][(\indexUser)]\in\realNumbers^{3\times1}$ is the location of the \ac{PD},  $\orientation[{\rm rx}][(\indexUser)]\in\realNumbers^{3\times1}$ is the orientation of the \ac{PD}, $\Ar$ is the area of \ac{PD} in ${\rm m}^2$, and $\FOV$ is the \ac{FOV} of \ac{PD}. {All orientation vectors are denoted with the letter $\textbf{q}$ and are unit vectors.}

\subsubsection{LOS Impulse Response}
The \ac{LOS} component of the channel impulse response between the source $\source$ and the receiver $\receiver$ is modeled by \cite{Barry_1993}
\begin{align}
\channelAtReflection[0][{t; \source, \receiver}] = \frac{\mode+1}{2\pi}\cos^\mode(\angleLOSandTX) \cos(\angleLOSandRX)\frac{\Ar}{\distance^2} \rect[{\frac{\angleLOSandRX}{\FOV}}]
\rect[{\frac{\angleLOSandTX}{\pi/2}}]\dirac[t-\propagationDelay]
~,
\label{eq:LOSchannel}
\end{align}
where $\angleLOSandTX$ is the angle between the source orientation vector $\orientation[{\rm tx}][(\indexLED)]$ and the incidence vector, $\angleLOSandRX$ is the angle between the receiver orientation vector $\orientation[{\rm rx}][(\indexUser)]$ and  the incidence vector, $\distance$ is the distance between the source and the receiver,  $\propagationDelay=\distance/\speedOfLight$ is the propagation delay, $\speedOfLight$ is the speed of light, $\dirac[\cdot]$ is the Dirac function, and $\rect[\cdot]$ is the rectangle function which takes the value 1 for $|x| \le 1$, and 0 otherwise.

While $\rect[{{\angleLOSandRX}/{\FOV}}]$ in \eqref{eq:LOSchannel} implies that the receiver can detect the light only when $\angleLOSandRX$ is less than $\FOV$, $\rect[{{\angleLOSandTX}/{(\pi/2)}}]$ ensures that the location of the receiver is in the \ac{FOV} of the source. The distance $\distance$ is the length of the incidence vector, which is given as $({\location[{\rm source}][(\indexLED)] - \location[{\rm receiver}][(\indexUser)]})$. The terms in \eqref{eq:LOSchannel} can be obtained as
$\cos(\angleLOSandTX)=\orientation[{\rm tx}][(\indexLED)\rm T]({\location[{\rm rx}][(\indexUser)] - \location[{\rm tx}][(\indexLED)]})/\distance,$
and $\cos(\angleLOSandRX)~=~-~\!\!\!\orientation[{\rm rx}][(\indexUser)\rm T]({\location[{\rm rx}][(\indexUser)] - \location[{\rm tx}][(\indexLED)] } )/\distance$. Some of the parameters are illustrated in Fig.~\ref{LambertianFigure}. 

\subsubsection{NLOS Impulse Response}
The \ac{NLOS} components of the channel between a LED and a \ac{PD} is obtained based on \emph{multiple-bounce impulse response} model described in \cite{Barry_1993}. In this model, light from a source $\source$ can reach a receiver $\receiver$ after infinite number of diffuse reflections and the channel impulse response is expressed as
\begin{align}
\channell[{t; \source, \receiver}] = \sum_{\degreeOfReflection=0}^{\infty}\channelAtReflection[\degreeOfReflection][{t; \source, \receiver}]~,
\label{eq:CIR}
\end{align}
where $t$ is the time index. Theoretically, $\channelAtReflection[\degreeOfReflection][{t; \source, \receiver}]$ can be expressed as a recursive function given by
\begin{align}
\channelAtReflection[\degreeOfReflection][{t; \source, \receiver}]=\int_{S} \reflectionCoefficient[{\rm ref}]\times \channelAtReflection[0][{t; \source, \{\location[{\rm ref}][], \orientation[{\rm ref}][], \deltaArea, \frac{\pi}{2}\}}] * \channelAtReflection[\degreeOfReflection-1][{t; \{\location[{\rm ref}][], \orientation[{\rm ref}][], 1\}, \receiver}] ,
\label{eq:CIRatReflection}
\end{align}
where $*$ denotes the convolution operation. In \eqref{eq:CIRatReflection}, the vector $\location[{\rm ref}][]\in\realNumbers^{3\times1}$ and the vector $\orientation[{\rm ref}][]\in\realNumbers^{3\times1}$ correspond to the location and the orientation of the reflector, respectively, $\deltaArea$ is the infinitesimal area of the reflector, and $\reflectionCoefficient[{\rm ref}]\in[0,1)$ is the reflection coefficient. In addition, the mode and the \ac{FOV} of the reflector are set to $1$ and $\pi/2$, respectively. {The real-valued DC channel gain \cite{haas_multireceiver,SDMA} between $\indexUser$th user and $\indexLED$th LED is then given by $\channel[\indexUser][\indexLED] = \int_{0}^{\infty} \channell[{t; \source, \receiver}] {\rm d}t.$

\begin{figure}[tb]
	\centering
	\includegraphics[width = 3.3 in]{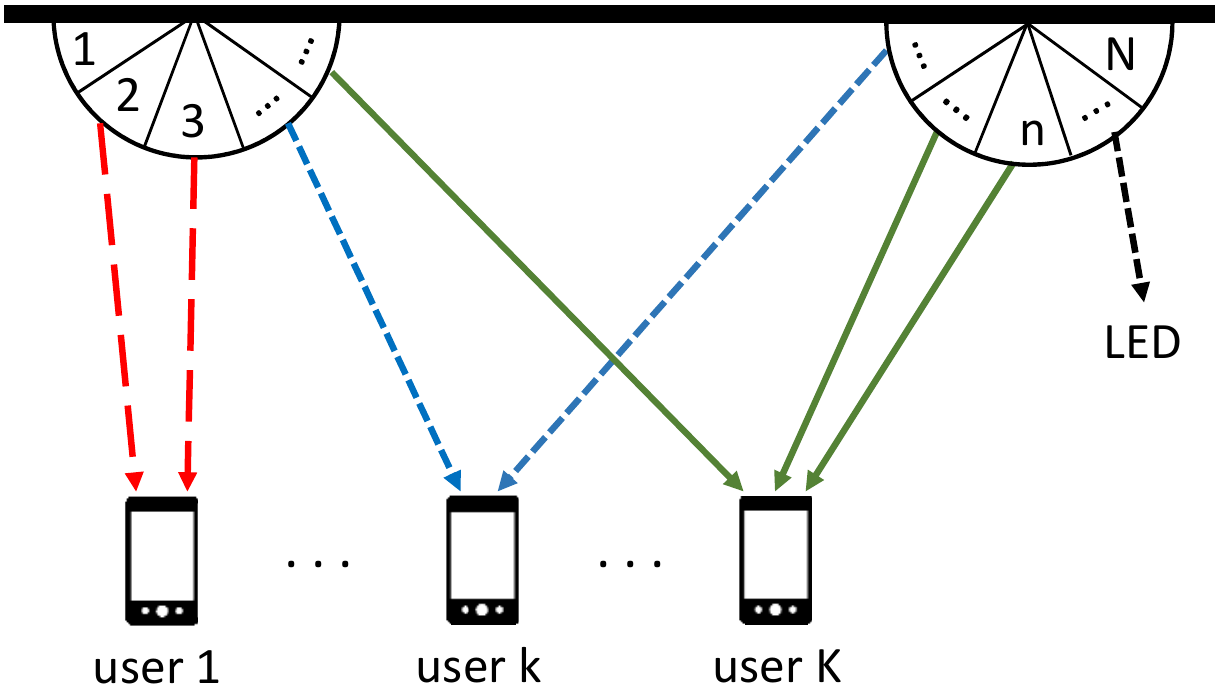}
	\caption{\small An example for LED assignment with $\NumberOfUsers$  users and $\NumberOfLEDs$ LEDs clustered at two VLC access points. All red dashed lines are assigned to user 1, all blue dotted lines are assigned to user $\indexUser$, while all solid green lines are assigned to user $\NumberOfUsers$.}
	\label{LED_assignment}
\end{figure}

\subsection{VLC Network and SINR Model}
We consider a VLC network as shown in Fig.~\ref{LED_assignment} where $\NumberOfUsers$ users are served by $\NumberOfLEDs$ LEDs such that $\NumberOfLEDs\gg\NumberOfUsers$. When more than a single LED is assigned to a user for a given case, the SINR of the $\indexUser$th user can be improved, and expressed as:
\begin{equation}
SINR{(\indexUser)} = \frac{\bigg(\displaystyle\sum_{\indexLED = 1}^{\NumberOfLEDs} {\PDresp\LEDConnectFlag[\indexUser][\indexLED] \channel[\indexUser][\indexLED]\power[\indexLED]}\bigg)^2}{N_0B + \displaystyle\sum_{\substack{\ell = 1 \\ \ell \neq k} }^{\NumberOfUsers}\bigg(\displaystyle\sum_{\indexLED = 1}^{\NumberOfLEDs} {\PDresp\LEDConnectFlag[\ell][\indexLED]\channel[\indexUser][\indexLED]\power[\indexLED]}\bigg)^2},
\label{SINR}
\end{equation}
where $\PDresp$ is the responsivity of the PD, $\power[\indexLED]$ is the standard deviation of the transmitted signal, $N_0$ is the spectral density of the additive white Gaussian noise (AWGN), and $B$ is the communication bandwidth. The connectivity variable is denoted by $\LEDConnectFlag[\indexUser][\indexLED]$, which is equal to $1$ when $\indexLED$th LED is assigned to $\indexUser$th user, and equal to $0$ otherwise:
\begin{align}
\LEDConnectFlag[\indexUser][\indexLED] \triangleq \begin{cases}
1 & \mbox{if \textit{\indexLED}th LED serves \textit{\indexUser}th user} \\
0 & \mbox{if \textit{\indexLED}th LED does not serve to \textit{\indexUser}th user}
\end{cases}~.
\end{align} 
 
To clarify our assumptions for the SINR in~\eqref{SINR} and provide further insights, two remarks are in order.

\begin{remark}
{\rm Several assumptions are made for the SINR in~\eqref{SINR} to hold. First, it is assumed that the transmission times of all LEDs are synchronized. Moreover, we assume that energy from the signals arriving to user-$k$ from the LEDs serving to that user can be coherently aggregated as in the numerator of~\eqref{SINR}. This may for example be possible considering a guard period among consecutive symbols~\cite{he2013m} where the delays from different LEDs can be absorbed, and inter-symbol-interference (ISI) effects~\cite{Nakagawa,biagi2013adaptive} are neglected. For modulation formats such as orthogonal frequency division multiplexing (OFDM) based VLC, delays of the signals arriving from different LEDs to a user can introduce phase changes at different subcarriers~\cite{wang2015multiuser}, which is not explicitly considered here, and their impact on the SINR is left as a future work. Finally, we also assume that the interference signals coming from a group of LEDs serving to the $\ell$th user are assumed to add up linearly at the desired user (for example, again considering a guard interval to absorb the energy). }
\end{remark}

\begin{figure}[tb]
	\centering
	\includegraphics[width = 3.4 in]{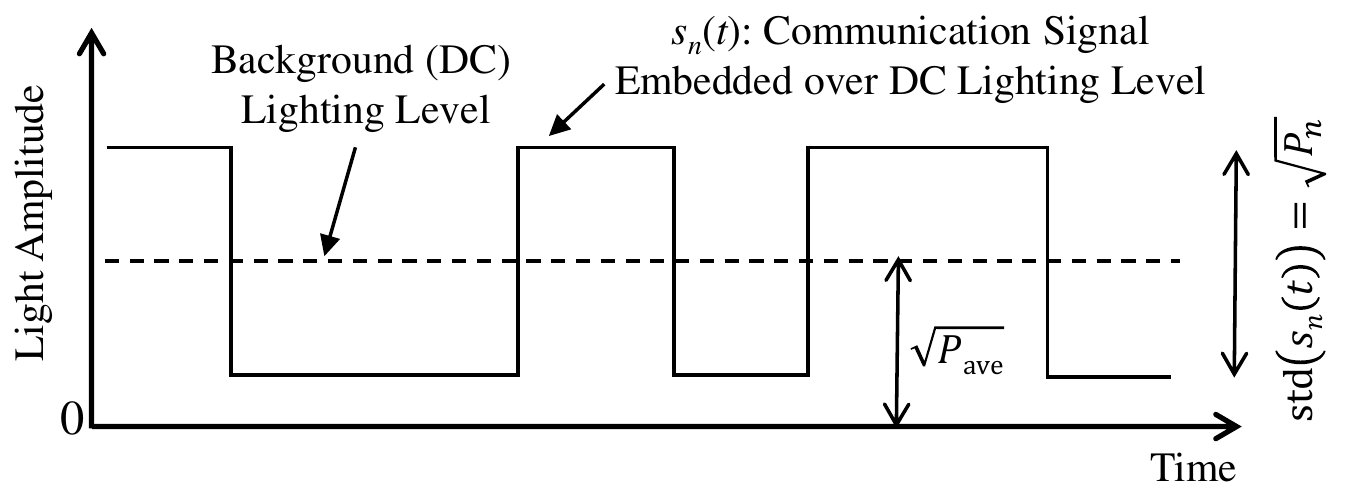}
	\caption{ Conceptual illustration for the background DC level and communication signal on $n$th LED for a transmitted VLC signal. }
	\label{Fig:ConceptPower}
\end{figure}

\begin{remark} {\rm As shown in the conceptual illustration of Fig.~\ref{Fig:ConceptPower}, we consider that the communication signal $s_n(t)$ for the $n$th LED is carried on the background DC light intensity (with power $P_{\rm ave}$) that is normally used for illumination. Zero mean communication signal $s_n(t)$ has a power $P_n = {\rm var}(s_n(t))$, and therefore a standard deviation of $p_n=\sqrt{P_n}={\rm std}(s_n(t))$. The $p_n$ will be referred in the rest of the paper as the \emph{power coefficient} for $n$th LED, and $\powerVector =[\power[1],...,\power[\NumberOfLEDs]]$ will be referred as the power coefficient vector.
We assume that all LEDs provide the same background light intensity ($\sqrt{P_{\rm ave}}$), which can be adjusted as desired, i.e., dimmed. On the other hand, we consider that $p_n$'s for different LEDs can be adjusted individually, considering $0\,{\le}\,p_n\,{\le}\,p_{\rm max}$ where $p_{\rm max}$ is the maximum power attainable based on the LED saturation output and $P_{\rm ave}$.}
\end{remark}

The numerator term of \eqref{SINR} characterizes the received total signal power from multiple LEDs serving user $\indexUser$. These LEDs transmit the same signal, 
and we sum all components to find the aggregate signal strength. The second term of the denominator represents the interference from LEDs serving simultaneously to $K-1$ other users. Likewise, the LEDs serving user $\ell~(\ell\neq k)$ transmit the same signal, therefore their sum is considered as one interference component. Finally, using $\LEDConnectFlag[\indexUser][\indexLED]$s, we can construct a connectivity matrix $\textbf{A}$ as follows:
\begin{align}
\textbf{A}=
\begin{bmatrix}
\LEDConnectFlag[1][1] & \LEDConnectFlag[1][2] & \dots & \LEDConnectFlag[1][\NumberOfLEDs] \\
\LEDConnectFlag[2][1] & \LEDConnectFlag[2][2] & \dots & \LEDConnectFlag[2][\NumberOfLEDs] \\
\vdots & \vdots & \ddots & \vdots \\
\LEDConnectFlag[\NumberOfUsers][1] & \LEDConnectFlag[\NumberOfUsers][2] & \dots & \LEDConnectFlag[\NumberOfUsers][\NumberOfLEDs]
\end{bmatrix},
\label{Eq:matrixA}
\end{align}
which can also capture the assignment of multiple LEDs to each user. {Note that only a single element in a column of $\textbf{A}$ can be one, and all other elements are zeros, since an LED is assumed to serve at most one user at a time. While techniques such as non-orthogonal multiple access (NOMA)~\cite{yin2016performance} and multi-user MIMO~\cite{wang2015multiuser} are available where one LED may serve to more than one user, they require higher complexity, and our main motivation in this paper is to take advantage of large number of directional LEDs for a simple yet efficient design}. On the other hand, sum of each row in~\eqref{Eq:matrixA} {is an integer greater or equal to one}, since multiple LEDs can serve a single user. 

One of our goals in this paper is to find the matrix $\textbf{A}$ under different optimization criteria to maximize the capacity of users considering different constraints, where the capacity of the $k$th user is given by
\begin{equation}
R_{\indexUser} = B{\rm log_2}\Big(1+SINR{(\indexUser)}\Big).
\label{Eq:rate}
\end{equation} 

{We only study the downlink VLC capacity in this paper, and assume that users may have uplink connectivity through an RF technology such as Wi-Fi. Moreover, we also assume that all LEDs can transmit data in synchronization. These requirements can be accomplished by using a central controller that controls all LEDs as well uplink RF reception from users. LEDs can be driven via Power over Ethernet (PoE) directly by the central controller. Even though connecting all LEDs to central controller (implying a star topology) may increase the installation cost due to longer cable size, plug and play simplicity of PoE can decrease the labor costs~\cite{7224733}. Apart from making synchronization of LEDs and control of downlink/uplink connections easier, another advantage for the considered backhaul system is to remove the requirement for separate processors for each VLC access point.}

\subsection{Efficient Calculation of the Channel Response}
In practice, the integration in \eqref{eq:CIRatReflection} can be evaluated by using the method of Riemann summation. The summation can be further simplified by exploiting the recursive structure of \eqref{eq:CIRatReflection}. However, when the operations are applied in time domain, the proposed method in \cite{Barry_1993} may still be time consuming  {since the number of paths grows exponentially with the number of reflections,} and limits the number of reflections taken into account for calculating~\eqref{eq:CIR}. In order to avoid this limitation, we calculate \eqref{eq:CIRatReflection} based on channel frequency response and corresponding matrix formulation, which allows its efficient calculation for any given reflection order.

Assume that $\numberOfReflectors$ reflectors are taken into account in order to model the multipath channel in the environment.  Such a spatial discretization leads to \eqref{eq:CIRatReflection}. Then, we let $\Cmatrix[\frequency]\in\realNumbers^{(\numberOfReflectors+1)\times(\numberOfReflectors+1)}$  be a matrix where the entry at $i$th row and $j$th column is
\begin{align}
\{\Cmatrix[\frequency]\}_{ij} = \channelAtReflection[0][{t; \source^{(j)}, \receiver^{(i)}}]e^{{\rm j}2\pi\propagationDelay_{ij} \frequency}~,
\end{align}
where $i$ is the receiver index, $j$ is the source index, $\receiver^{(i)}$ is the $i$th receiver, $\source^{(j)}$ is the $j$th source, $\propagationDelay_{ij}$ is the propagation delay, and $f$ is the frequency index. Without loss of generality, $\receiver^{(1)}$ and $\source^{(1)}$ are the parameter sets for the LED and the \ac{PD}, respectively, and the reflectors are indexed by $i,j\in\{2,3,\dots,\numberOfReflectors+1\}$.
Then, the received signal components at the receiver and at the reflectors due to the light experiencing exactly $\degreeOfReflection>1$ bounces is obtained as
\begin{align}
\pVector_{\degreeOfReflection}(\frequency) = \big[\Cmatrix[\frequency]\Dmatrix\big]^{\degreeOfReflection} \pVector_{0}(\frequency)~,
\label{eq:recSignalFrequency}
\end{align}
where $\pVector_{\degreeOfReflection}(\frequency)$ is the received signal power due to the $\degreeOfReflection$th order reflections, $\Dmatrix$ is a diagonal matrix where the diagonal elements include the reflection coefficients with entry $\{\Dmatrix\}_{11}$ is set to 0, and the vector $\pVector_{0}(\frequency)$ consists of the received signal power at the nodes due to the \ac{LOS} component of the channel. The vector $\pVector_{0}(\frequency)$  can be calculated as $\pVector_{0}(\frequency) = \Cmatrix[\frequency] \initialVector~,$
where $\initialVector\in\realNumbers^{(\numberOfReflectors+1)\times 1}$ is the excitation vector with the first entry being $1$ and the rest of the entries being zeros.

By using the eigenvalue decomposition of $\Cmatrix[\frequency]\Dmatrix$, \eqref{eq:recSignalFrequency} can be rewritten as
\begin{align}
\pVector_{\degreeOfReflection}(\frequency) = \eigenMatrix \big[\eigenDiagonal(\frequency)\big]^{\degreeOfReflection} \eigenMatrix^{-1}\pVector_{0}(\frequency)~,
\label{eq:13eq}
\end{align}
where $\eigenMatrix\in\realNumbers^{(\numberOfReflectors+1)\times(\numberOfReflectors+1)}$ is a matrix whose columns are the eigenvectors of $\Cmatrix[\frequency]\Dmatrix$ and $\eigenDiagonal$ is the diagonal matrix whose diagonal elements are the corresponding eigenvalues. Therefore, \eqref{eq:recSignalFrequency} can be calculated efficiently for an arbitrary number of multipath reflections, since $\eigenDiagonal$ is a diagonal matrix. By evaluating \eqref{eq:13eq} for different frequencies, channel frequency response (CFR) can be calculated for a given bandwidth and one can calculate the channel impulse response (CIR) from CFR by using inverse Fourier transformation. 

{Note that the authors in \cite{1231653} also suggest an efficient CIR calculation method for MIMO systems by  decomposing the channel response to three different matrices which represent the receiver parameters, transmitter parameters, and indoor environment. However, \eqref{eq:13eq} is different than the model given in~\cite{1231653} as it takes the exact propagation delays into account, and simultaneously calculates the impact of all of the paths on the CFR for a given frequency.} The equation~\eqref{eq:13eq} is used in all simulations in Section~VI to generate multipath realizations. 


\section{LED Assignment to Users}\label{Sec:LED_Assign}
In this section, we consider the problem of LED assignment to users for a network as in Fig.~\ref{LED_assignment} considering two different scenarios. In the first scenario studied in Section~\ref{Sec:NoQoS}, there are no QoS guarantees and the network provides the highest sum rate or the highest proportionally fair sum rate. LED power control for this first scenario is also investigated in Section~\ref{Sec:PowerControl}. In the second scenario in Section~\ref{Sec:QoSGuarantees}, there are QoS guarantees and all users get  data rates proportional to their QoS ratios.

\subsection{LED Assignment without QoS Guarantees}\label{Sec:NoQoS}
When there are no QoS guarantees, the problem of maximizing throughput can be expressed as
\begin{equation}
\begin{split}
\big[\textbf{A}',\powerVector'\big] =~&{\rm arg~} \underset{\textbf{A},\powerVector}{\rm max}~~ \displaystyle\sum_{\indexUser=1}^{\NumberOfUsers} R_{\indexUser},\\
{\rm subject~to}~&0 \le \power[\indexLED] \le p_{\rm max} ~\forall \indexLED,\\
& \LEDConnectFlag[\indexUser][\indexLED] \in \{0,1\}, \\
&\displaystyle\sum_{\indexUser=1}^{\NumberOfUsers}{\LEDConnectFlag[\indexUser][\indexLED]} \leq 1 ~~\forall \indexLED,
\end{split}
\label{maxthroughput}
\end{equation}
where $\textbf{A}'$ and $\powerVector'$ are the connectivity matrix and the {power coefficient} vector which maximize the sum capacity over all possible values of $\textbf{A}$ and $\powerVector$, respectively. 

While the solution of \eqref{maxthroughput} maximizes the total throughput, it does not consider the fairness among the users, and it will assign most of the LEDs to the users with good SINRs. It has been shown that maximizing the total logarithmic throughput achieves proportional fairness \cite{3010473, 1498527}. If we aim at providing proportional fairness among users, the problem can be modified as follows:
\begin{equation}
\begin{split}
\big[\textbf{A}',\powerVector'\big] =~&{\rm arg~} \underset{\textbf{A},\powerVector}{\rm max}~~ \displaystyle\sum_{\indexUser=1}^{\NumberOfUsers} {\rm log}( R_{\indexUser}),
\end{split}
\label{logthroughput}
\end{equation}
subject to the same constraints as \eqref{maxthroughput}. Solution of this problem maximizes the throughput while distributing LEDs in a proportionally fair manner with respect to users' channel conditions. This solution is also a special case of QoS-oriented LED assignment to be studied in Section~\ref{Sec:QoSGuarantees}.

The optimization problems in \eqref{maxthroughput} and \eqref{logthroughput} are generally hard to solve since they include both discrete and continuous variables. Such problems are called \textit{binary mixed-integer programming problems} in the literature \cite{NAV3800210404}. {To simplify the problem, we follow a suboptimal approach and solve the problem in two steps. We first assume identical power level to all LEDs, which is the maximum power coefficient $p_{\rm max}$, and solve the LED assignment problem alone. In the second step, we optimize the powers coefficients of LEDs, which is addressed in the next subsection of this paper.}

In \cite{4146798}, the problem of assigning users to base stations for 3G networks considering proportional fairness is shown to be an NP-hard problem, which means there is no algorithm that can find the optimum solution in polynomial time~\cite{Woeginger2003}. The problem in \cite{4146798} is not the same as \eqref{logthroughput}; however if we switch the role of users and LEDs, we can establish a connection between the two problems. In 3G networks, there are large number of users and fewer number of base stations, and one user is connected to one base station at a given time. In our model, we assume that there are larger number of LEDs than users, and an LED will be assigned to a single user at a time. It can be shown that the problem in \cite{4146798} can be reduced to \eqref{logthroughput}, hence, assigning LEDs to users is also an NP-hard problem.

NP-hard problems can be solved via exhaustive search; however, when the number of elements (LEDs, users) increase, running time for exhaustive search becomes very large. In Section~\ref{ComplexitySect}, we present computational complexity of the exhaustive search along with the proposed assignment algorithms, and show that the exhaustive search is computationally infeasible. Even though one may find more efficient solutions than the exhaustive search, it is not possible to find a solution with a running time proportional to a polynomial function of the problem size. Thus, we focus on developing two different heuristic techniques that find close-to-optimal solutions at low computational complexity. We also compare the performance of these heuristics with exhaustive search for small number of users/LEDs in Section~VII.

\subsubsection{Highest RSS based assignment (HRS)}
In this algorithm, we assign an LED to the user that receives the highest RSS from that LED. Considering that all LEDs provide the same {transmit power} and all users have the same PD responsivity, the RSS is proportional to the channel gain. Therefore, the user who will be served by the $\indexLED$th LED is given by
\begin{equation}
\hat{\indexUser} = {\rm arg~} \underset{\indexUser}{\rm max}~\channel[\indexUser][\indexLED],
\end{equation}
and hence, we have $\LEDConnectFlag[\hat{\indexUser}][\indexLED]=1$ for the particular LED. For all the other $\indexUser\neq \hat{\indexUser}$, we have $\LEDConnectFlag[\indexUser][\indexLED]=0$. All LEDs are assigned in the same way without explicitly considering the fairness among users. This is the simplest algorithm and gives high sum throughput, and we will refer to this algorithm as the HRS algorithm.

\subsubsection{Weighted signal strength based assignment (WSS)}
In this algorithm, we scale RSS with the inverse of the total received signal power by that user. In particular, the weighted RSS for $\indexLED$th LED by $\indexUser$th user is given by
\begin{equation}
\Psi_{\indexUser\indexLED} = \frac{\PDresp\power[\indexLED]\channel[\indexUser][\indexLED]} {\displaystyle\sum_{m = 1}^{\NumberOfLEDs} {(\PDresp\power[m]\channel[\indexUser][m])^2}} \propto \frac{\channel[\indexUser][\indexLED]} {\displaystyle\sum_{m = 1}^{\NumberOfLEDs} {\channel[\indexUser][m]^2}}.
\label{WSS}
\end{equation}
The proportionality in \eqref{WSS} holds because {power coefficients of the LEDs are assumed to be identical at this stage of the problem.} Afterwards, each LED is assigned to the user with highest weighted signal strength. In other words, for the $\indexLED$th LED, the user who will be served by that LED is given by
\begin{equation}
\hat{\indexUser} = {\rm arg~} \underset{\indexUser}{\rm max}~\Psi_{\indexUser\indexLED},
\end{equation}
and hence, we have $\LEDConnectFlag[\hat{\indexUser}][\indexLED]=1$. For all the other $\indexUser\neq \hat{\indexUser}$ and LED $\indexLED$, we have $\LEDConnectFlag[\indexUser][\indexLED]=0$. This method is expected to provide a more fair allocation than just assigning each LED to the user which receives highest RSS from that LED. That is because it takes into account the aggregate signal power a user receives and gives priority to users with low overall RSS.

\subsection{LED Power Control}\label{Sec:PowerControl}
In this subsection, assuming that LEDs are assigned using an approach as discussed earlier, we consider the problem of power control over the assigned LEDs to all users. In order to solve the optimization problem in either \eqref{maxthroughput} or \eqref{logthroughput} to find the optimal power coefficients $p_n$, we formulate the respective Lagrange dual function as follows
\begin{align}\label{eq:lagrange_func} 
\mathcal{L}(\powerVector, \boldsymbol \lambda) &{=} \displaystyle\sum_{\indexUser=1}^{\NumberOfUsers} \tilde{R}_{\indexUser}\,{+}\displaystyle\sum_{\indexLED = 1}^{\NumberOfLEDs}\lambda_n(p_n{-}\,p_{\rm max}) \,{-}\displaystyle\sum_{\indexLED = 1}^{\NumberOfLEDs}\lambda_{n{+}N}p_n \,, 
\end{align}
where $\tilde{R}_{\indexUser}$ is the generic rate function given as 
\begin{align}\label{eq:generic_rate}
\tilde{R}_{\indexUser}\,{=}\,
\begin{cases}
R_k \,, & \text{for the optimization in \eqref{maxthroughput}} \\
\log (R_k) \,, & \text{for the optimization in \eqref{logthroughput}}
\end{cases}~,
\end{align}
and all the Lagrange multipliers ($\lambda_n$'s) are stacked in the vector $\boldsymbol \lambda$. The optimal power coefficient can be solved either by i) finding roots of the derivative of the Lagrange function in \eqref{eq:lagrange_func} with respect to unknowns $\bold{p}$ and $\boldsymbol \lambda$ via Newton based methods, or ii) directly minimizing the optimization problem in \eqref{maxthroughput} or \eqref{logthroughput} by interior-point or trust-region methods~\cite{Boyd04ConOpt}. For either approach, we need to derive the Jacobian and the Hessian matrices which will be provided in Corollary~\ref{theo:jacob_hessian}. Before that, we first give the first and second-order derivatives of the rate functions in Theorem~\ref{theo:rate_derivatives}, which will be necessary for the derivation in Corollary~\ref{theo:jacob_hessian}.

\begin{theorem}\label{theo:rate_derivatives}
Defining $f(\cdot)$ to be the mapping function for the LED assignment scheme such that $\ell\,{=}\,f(m)$ is the index for the user served by the $m$th LED, the first-order derivative of the rate $R_k$ with respect to the power coefficient is given as
\begin{align}
\frac{\partial R_{\indexUser}}{\partial p_m} = \frac{B}{\rm ln2}\frac{2 S_{\ell k}\channel[\indexUser][m]}{T_\indexUser} \, C_\indexUser^m,
\label{eq:rate_derivative_1st}
\end{align}
where $S_{\ell k}{=}\sum_{\indexLED=1}^{\NumberOfLEDs}\LEDConnectFlag[\ell][\indexLED]\channel[\indexUser][\indexLED]p_n$, $T_\indexUser\,{=}\,N_0B/r\,{+}\,\sum_{\ell=1}^\NumberOfUsers S_{\ell k}^2$, and $C_\indexUser^m{=}\delta(k,l){-}SINR(\indexUser)(1{-}\delta(k,l))$ with $\delta (\cdot,\cdot)$ being the Kronecker delta function. Similarly, the second-order derivative of the rate is given as
\begin{align}\label{eq:rate_derivative_2nd}
\frac{\partial^2 R_{\indexUser}}{\partial p_m \partial p_n} = \frac{B}{\rm ln2}\frac{4 S_{\ell k} h_{km} S_{\ell'k} h_{kn}}{T_k^2} \, E_k^{m,n},
\end{align}
where $\ell'\,{=}\,f(n)$, and,
\begin{align}
E_k^{m,n} {=} \begin{cases}
\displaystyle {-}1 + \frac{T_k}{2S_{\ell k}^2}\delta(\ell,\ell') \,, & k {=} \ell \\
\displaystyle SINR(k)\left( 2\,{+}\,SINR(k)\,{-}\,\frac{T_k}{2S_{\ell k}^2}\delta(\ell,\ell')\right) \,, & k {\neq} \ell\\
\end{cases}.\nonumber 
\end{align}
While the derivatives of the generic rate function $\tilde{R}_{\indexUser}$ in \eqref{eq:generic_rate} is given directly by \eqref{eq:rate_derivative_1st} and \eqref{eq:rate_derivative_2nd} for the optimization problem in \eqref{maxthroughput}, respective derivatives for the optimization problem in \eqref{logthroughput} are given as
\begin{align*}\label{eq:rate_derivatives_logthrouput}
\frac{\partial \tilde{R}_{\indexUser}}{\partial p_m}&\,{=}\, \frac{1}{R_k}\frac{\partial R_{\indexUser}}{\partial p_m}\,, \;\;
\frac{\partial^2 \tilde{R}_{\indexUser}}{\partial p_m \partial p_n} \,{=}\,\frac{1}{R_k^2}\frac{\partial R_{\indexUser}}{\partial p_m}\frac{\partial R_{\indexUser}}{\partial p_n} {+} \frac{1}{R_k}\frac{\partial^2 R_{\indexUser}}{\partial p_m \partial p_n}
\end{align*}
\end{theorem}
\begin{IEEEproof}
See Appendix~\ref{app:rate_derivatives}.
\end{IEEEproof}

\begin{corollary}\label{theo:jacob_hessian}
The Jacobian of the Lagrange dual function in \eqref{eq:lagrange_func} is given as
\begin{align}
J_m &\,{=}\,\frac{\partial \mathcal{L}(\powerVector, \boldsymbol \lambda)}{\partial p_m}\,{=}\,\displaystyle \sum_{\indexUser = 1}^{\NumberOfUsers}\frac{\partial \tilde{R}_{\indexUser}}{\partial p_m}\,{+}\,\lambda_m\,{-}\,\lambda_{m{+}\NumberOfLEDs},\nonumber
\end{align}
for $1 \leq m \leq N$, and,
\begin{align}
J_m &\,{=}\,\frac{\partial \mathcal{L}(\powerVector, \boldsymbol \lambda)}{\partial \lambda_{m{-}N}} 
\,{=}\begin{cases}
p_{m{-}N}\,{-}\,p_{\rm max}\,, & \!\!\! N{+}1 \leq m \leq 2N \\
-p_{m{-}2N}\,, 				   & \!\!\! 2N{+}1 \leq m \leq 3N
\end{cases}.\nonumber
\end{align}
Similarly, the Hessian of \eqref{eq:lagrange_func} is given as
\begin{align}
G_{m,n} &\,{=}\, \frac{\partial J_m}{\partial p_n} {=} \begin{cases}
\displaystyle\sum_{\indexUser=1}^{\NumberOfUsers}\frac{\partial^2 \tilde{R}_{\indexUser}}{\partial p_m \partial p_n}, & \!\!\! 1 \leq m \leq N \\
\delta(m{-}N,n), & \!\!\! N{+}1 \leq m \leq 2N \\
{-}\,\delta(m{-}2N,n), & \!\!\! 2N{+}1 \leq m \leq 3N
\end{cases},\nonumber
\end{align}
for $1 \leq n \leq N$, and,
\begin{align}
G_{m,n} \,{=}\, \frac{\partial J_m}{\partial \lambda_{n{-}N}} \,{=}\, \delta(m,n{-}N){-}\delta(m{+}\NumberOfLEDs,n{-}N),\nonumber
\end{align}
for $1 \leq m \leq N, N{+}1 \leq n \leq 3N$, and $0$ otherwise.
\end{corollary}
\begin{IEEEproof}
The Jacobian and the Hessian are the first and the second-order derivatives of \eqref{eq:lagrange_func}, which can be readily computed using the derivatives in Theorem~\ref{theo:rate_derivatives}.
\end{IEEEproof}

\subsection{LED Assignment with QoS Guarantees}\label{Sec:QoSGuarantees}
To make sure all users are allocated sufficient resources to satisfy their QoS, in this section we also study the LED assignment problem with predetermined QoS ratios among the users. {Due to the complexity of the problem and space limitations, we leave the power control for the LED assignment with QoS constraints as a future study.}  QoS guarantees enable users with higher priorities to receive higher data rates by proportionally allocating the resources to the users based on their QoS requirements. When QoS ratios are provided, the problem in \eqref{logthroughput} can be modified as follows
\begin{equation}
\begin{split}
\textbf{A}' =~&{\rm arg~} \underset{\textbf{A}}{\rm max}~~ \displaystyle\sum_{\indexUser=1}^{\NumberOfUsers} R_{\indexUser},\\
{\rm subject~to}~&\LEDConnectFlag[\indexUser][\indexLED] \in \{0,1\}, \\
&\displaystyle\sum_{\indexUser=1}^{\NumberOfUsers}{\LEDConnectFlag[\indexUser][\indexLED]} \leq 1 ~~\forall \indexLED,\\ 
& \frac{R_1}{\nu_1} = \frac{R_2}{\nu_2} = ... = \frac{R_\NumberOfUsers}{\nu_\NumberOfUsers},
\end{split}
\label{propAssign}
\end{equation}
where $\nu_\indexUser$ is the QoS ratio for user $\indexUser$. Assigning all users the same QoS ratios means maximizing the sum rate while making all $R_\indexUser$'s equal and it corresponds to max-min problem, which is maximizing the rate of the minimum rate user \cite{shen2005adaptive}. Therefore, max-min problem is a special case of the problem in \eqref{propAssign}. The problem has similarities with the subchannel block allocation of multiuser OFDM systems~\cite{rhee2000increase}, assuming frequency selective quasistatic channels where channels do not vary within a block of transmission. However, while assigning different number of LEDs to the users alters the SINR of the users, assigning different number of subchannels to the the users changes the assigned bandwidth to the users.

As a solution to problem \eqref{propAssign}, we present a new LED assignment heuristic given in Algorithm~1. We define $\delta_{\indexUser,\indexLED}~\!\triangleq~\!(\power [\indexLED]\channel[\indexUser][\indexLED])^2/N_0$ as the signal-to-noise ratio (SNR) for user $\indexUser$ and LED $\indexLED$, $\assignedto$ as the set of LEDs assigned to user $\indexUser$, and $\unassignedLEDs$ as the set of unassigned LEDs. The algorithm initially assigns one LED to each user, which is chosen based on the highest SNR to that user. Then, iteratively, the algorithm lets the user with the least proportional capacity to pick up an LED. The user picks up the LED that provides highest SNR from the available LEDs ($\unassignedLEDs$), and the algorithm iterates until all LEDs are assigned. \\

\begin{algorithm}[tb]
	\caption{Proportional Rate Algorithm}
	\begin{algorithmic}[1]
		\STATE Initialize, $R_\indexUser = 0$, $\assignedto = 0$ for $\indexUser = 1,2,...,\NumberOfUsers$ and $\unassignedLEDs=\{1,2,...,\NumberOfLEDs\}$
		\STATE For $\indexUser = 1$ to $\NumberOfUsers$
		\vspace{1mm}
		\Indent
		a) find $\indexLED$ providing $ \delta_{\indexUser,\indexLED} \ge \delta_{\indexUser,j}$ $~\forall j \in \unassignedLEDs$
		\EndIndent
		\Indent
		b) set $\assignedto$ = $\assignedto \cup \{\indexLED\},~  \unassignedLEDs = \unassignedLEDs - \{\indexLED\}$, update $R_{\indexUser}$ using \eqref{Eq:rate}
		\EndIndent
		\STATE While $\unassignedLEDs\neq O$
		\vspace{1mm}
		\Indent
		a) find $\indexUser$ providing $R_{\indexUser}/\nu_{\indexUser} \le R_i/\nu_i$, for all $i$, $1 \le i \le \NumberOfUsers$
		\EndIndent
		\Indent
		b) for the found $\indexUser$, find $\indexLED$ providing $\delta_{\indexUser,\indexLED} \ge \delta_{\indexUser,j}$, $~\forall j \in \unassignedLEDs$
		\EndIndent
		\Indent
		c) for the found $\indexUser$ and $\indexLED$, set $\assignedto = \assignedto \cup \{\indexLED\}$,~ $\unassignedLEDs = \unassignedLEDs - \{\indexLED\}$ and update $R_\indexUser$ using \eqref{Eq:rate}
		\EndIndent
	\end{algorithmic}
	\label{alg}
\end{algorithm}

\subsection{{Remarks on LED Assignment Protocol}}
{In order to do the LED assignment, the central controller needs to learn the RSS at each user observed from each LED. This information can be measured at each user and reported to the central controller using the uplink RF channel. While there may be  different protocols to achieve this, we will provide one example based on the LTE technology~\cite{sesia2011lte}. First, similar to LTE, periodically transmitted downlink synchronization/discovery sequences can uniquely characterize the LED identity. For example, Zadoff-Chu and $m$-sequences in LTE~\cite[Ch.~7]{sesia2011lte} can uniquely identify 504 difference base stations due to their excellent correlation characteristics. }

{Subsequently, for each identified LED, users can measure the RSS from that LED over some orthogonal pilot symbols~\cite[Ch.~8]{sesia2011lte} (or even the synchronization sequences themselves~\cite[Ch.~7]{sesia2011lte}). If the RSS measurements at a user trigger some measurement reporting criteria (e.g., RSS from serving LEDs falling below a threshold for some duration~\cite[Ch.~3]{sesia2011lte}), measurements from neighboring LEDs can be reported to the central controller. Alternatively, measurements may also be reported periodically, which  are then used for updating the LED assignments considering also the measurements from other users. For LED assignment with QoS guarantees, such measurement reports may also include $\nu_k$ for user $k$ if there are any changes in the QoS requirements of users. Addressing unique challenges/aspects for implementing such a protocol for the multi-LED VLC framework specifically considered in this paper is left as a future study.}

\begin{figure}[tb]
	\centering
	\includegraphics[width = 3.4 in]{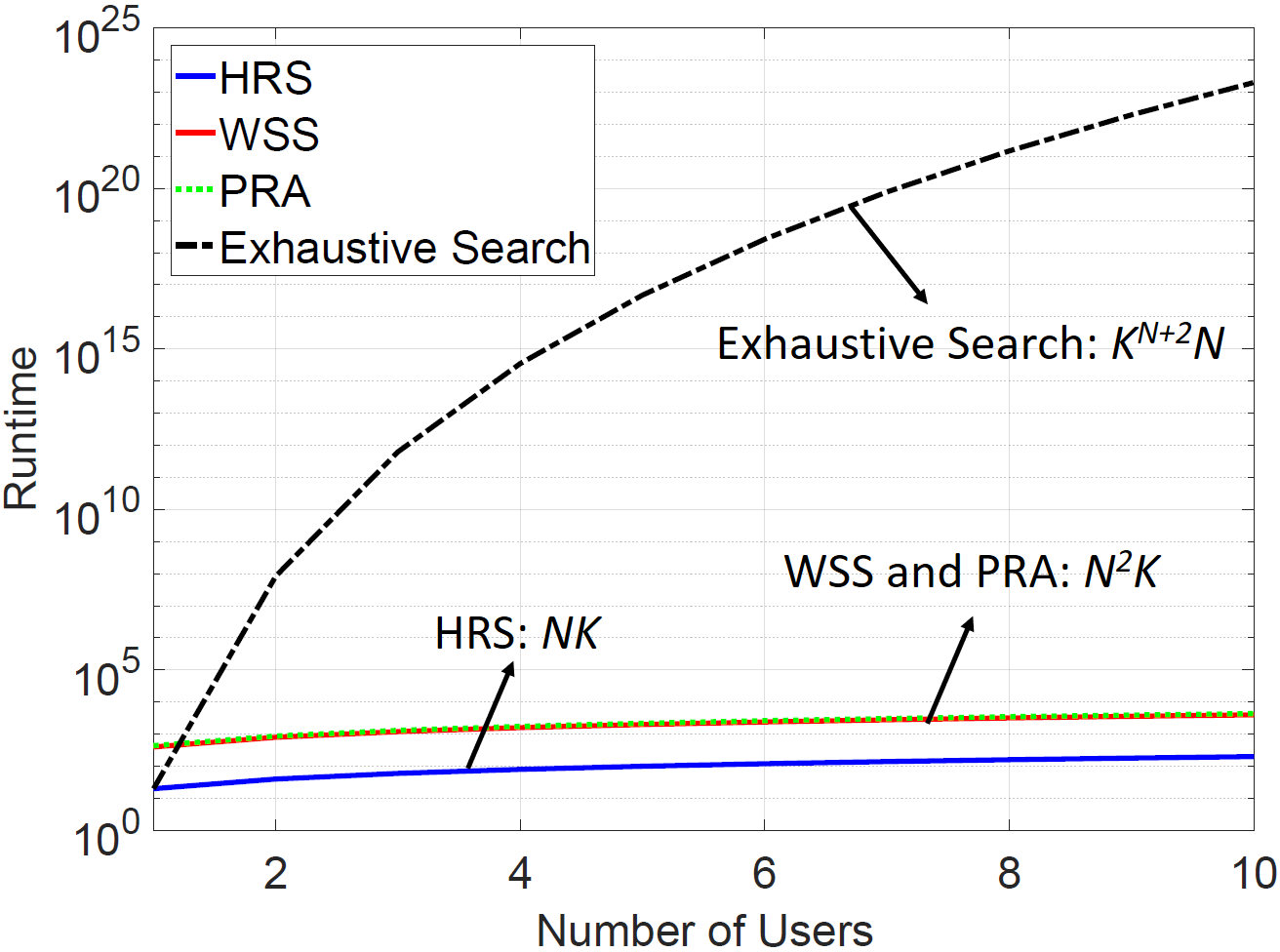}
	\caption{Comparison of time complexities of different algorithms for $\NumberOfLEDs = 20$ on logarithmic scale.}
	\label{AlgorithmComplexity}
	\vspace{-3mm}
\end{figure}

\subsection{Computational Complexity for LED Assignment}
\label{ComplexitySect}
In this section, we provide some remarks on the computational complexities for LED assignment techniques with and without QoS guarantees. First, the running time to find overall optimal LED assignment by exhaustive search is $O(\NumberOfUsers^{\NumberOfLEDs+2}\times\NumberOfLEDs)$. The reason for that is, there are $\NumberOfUsers^{\NumberOfLEDs}$ different possibilities to assign LEDs to users, which requires $\NumberOfUsers^{\NumberOfLEDs}$ iterations. For each iteration, the data rate needs to be calculated for $\NumberOfUsers$ users, and each rate calculation requires SINR computation as given in~\eqref{SINR}. The most complex operation in~\eqref{SINR} is the calculation of interference that takes $\NumberOfLEDs\times\NumberOfUsers$ iterations, and the total running time of the algorithm becomes proportional to $\NumberOfUsers^{\NumberOfLEDs+2}\times\NumberOfLEDs$. The running time of different algorithms are compared in Fig.~\ref{AlgorithmComplexity}, which confirms that the exhaustive search is computationally infeasible even for few number of users.

Second, the running time of the HRS algorithm is $O(\NumberOfLEDs \times \NumberOfUsers)$, because the time to find the maximum RSS value for an LED observed at $\NumberOfUsers$ different users is proportional to $\NumberOfUsers$, which is repeated $\NumberOfLEDs$ times by the number of LEDs. Running time for the WSS algorithm is $O(\NumberOfLEDs^2 \times \NumberOfUsers)$. The required time for the calculation of weighted signal strength as in \eqref{WSS} is proportional to the number of LEDs, i.e., $\NumberOfLEDs$ (assuming the squares and the division are constant operations), and it will be performed for $\NumberOfLEDs$ LEDs and $\NumberOfUsers$ users, which takes time proportional to $\NumberOfLEDs^2 \times \NumberOfUsers$. Then, choosing the maximum weighted strength from  $\NumberOfUsers$ users and repeating it for $\NumberOfLEDs$ LEDs also takes time proportional to $\NumberOfLEDs \times \NumberOfUsers$, which may be considered a smaller order function and insignificant.

Finally, the running time of the LED assignment with QoS guarantees in Section~\ref{Sec:QoSGuarantees} is $O(\NumberOfLEDs^2 \times \NumberOfUsers)$. There is a while loop and a for loop in the algorithm, and total number of iterations in both loops is $\NumberOfLEDs$. In both loops, the most complicated operation is updating the rate of a user, which requires SINR calculation that takes time proportional to $\NumberOfLEDs\times\NumberOfUsers$. It is executed once in any of $\NumberOfLEDs$ iterations, so the total running time is proportional to $\NumberOfLEDs^2\times\NumberOfUsers$ (See Fig.~\ref{AlgorithmComplexity}).

\section{Diversity Combining} \label{Sec:Diversity_Comb}

In Section~\ref{Sec:LED_Assign} we discussed the LED assignment problem, presented our solutions, demonstrated their time complexities, and {provided a power control approach for improved performance.} In this section, we propose an advanced receiver combining method that further improves the SINR, by taking advantage of the LED assignment information to the users. When multiple PDs are used at a receiver, the SINR of $\indexUser$th user after combining over multiple PDs can be calculated as:
\begin{equation}
\begin{split}
SINR{(\indexUser)} = \frac{\bigg(\displaystyle\sum_{\indexLED = 1}^{\NumberOfLEDs}\displaystyle\sum_{\indexPD=1}^{\NumberOfPDs} \PDresp\LEDConnectFlag[\indexUser][\indexLED]\power[\indexLED]w_{\indexPD}\channel[\indexUser(\indexPD)][\indexLED]\bigg)^2}{\displaystyle\sum_{\indexPD=1}^{\NumberOfPDs}w_{\indexPD}^2N_0B + \displaystyle\sum_{\substack{\ell = 1 \\ \ell \neq k}   }^{\NumberOfUsers}\bigg(\displaystyle\sum_{\indexLED = 1}^{\NumberOfLEDs}\displaystyle\sum_{\indexPD=1}^{\NumberOfPDs}\PDresp\LEDConnectFlag[\ell][\indexLED] \power[\indexLED]w_{\indexPD}\channel[\indexUser(\indexPD)][\indexLED]\bigg)^2},
\label{SINR_multi}
\end{split}
\end{equation}
where $w_{\indexPD}$ is the weight for $\indexPD$th PD, $\NumberOfPDs$ is the number of PDs on a receiver, $\channel [\indexUser(\indexPD)][\indexLED]$ is the channel attenuation between $\indexLED$th LED and $\indexPD$th PD of $\indexUser$th user, and the second term in the denominator represents the sum of all interference signal powers at all PDs from all LEDs excluding the LED group which serves the $\indexUser$th user. Choosing of $w_{\indexPD}$ for combining signals at the receiver can be achieved using the MRC or the OC approaches, as will be discussed next.

\subsection {Maximum Ratio Combining}
The MRC uses the signals received from different PDs with a proportional weight to the SINR observed at each PD. Weight of $\indexPD$th PD is calculated as
\begin{equation}
w_{\indexPD} = \frac{\bigg(\displaystyle\sum_{\indexLED = 1}^{\NumberOfLEDs} \PDresp\LEDConnectFlag[\indexUser][\indexLED]\power[\indexLED]\channel[\indexUser(\indexPD)][\indexLED]\bigg)^2}{N_0B + \displaystyle\sum_{\substack{\ell = 1 \\ \ell \neq k}   }^{\NumberOfUsers}\bigg(\displaystyle\sum_{\indexLED = 1}^{\NumberOfLEDs}\PDresp\LEDConnectFlag[\ell][\indexLED] \power[\indexLED]\channel[\indexUser(\indexPD)][\indexLED]\bigg)^2}.
\label{MRC_weight}
\end{equation}
The numerator of \eqref{MRC_weight} is for the received signal power at $\indexPD$th PD, and the denominator is for the sum of noise and interference. The MRC is a heuristic to use all data with a proportional ratio to maximize the combined SINR, however, it assumes that the signals received at different PDs are uncorrelated. While this approach is successful at suppressing the white noise, it yields suboptimal performance for correlated noise or interference. 

\subsection {OC with Unknown Grouping Information}
The OC provides higher SINR performance than the MRC by suppressing correlated interference. In order to calculate the weights of the $\indexUser$th user, denote
\begin{equation} 
\ReceivedPower [\indexUser][\indexPD][\indexUser] = \PDresp\sum_{\indexLED = 1}^{\NumberOfLEDs} {\LEDConnectFlag[\indexUser][\indexLED]\power[\indexLED]\channel[\indexUser(\indexPD)][\indexLED]},
\end{equation}
to be the sum of the received desired signals at $\indexUser$th user's $\indexPD$th PD. We can build a vector 
\begin{equation} 
\signalVector[k][k] = [\ReceivedPower [\indexUser][1][\indexUser], \ReceivedPower [\indexUser][2][\indexUser], ... , \ReceivedPower[\indexUser][\NumberOfPDs][\indexUser] ]^T
\end{equation}
which includes the received desired signals through different PDs. The weighting vector for OC can then be calculated as
\begin{equation}
\textbf{w}=\textbf{R}^{-1}\signalVector[k][k],
\end{equation} 
where $\textbf{R}$ is the interference-plus-noise correlation matrix of the received signal, explicitly given by 
\begin{align}
\textbf{R} = N_0B\textbf{I} + \displaystyle\sum_{\indexLED = 1}^{\NumberOfLEDs}{(1-\LEDConnectFlag[\indexUser][\indexLED])E[\channelVector[\indexUser][\indexLED]\channelVector[\indexUser][\indexLED]^T]},
\label{intCorr}
\end{align}
where 
\begin{align}
\channelVector[\indexUser][\indexLED] = \PDresp\power [\indexLED][\channel[\indexUser(1)][\indexLED], \channel [\indexUser(2)][\indexLED], ... , \channel [\indexUser(\NumberOfPDs)][\indexLED]]^T,
\end{align} 
which is the signal vector that the $\indexUser$th user captures from the $\indexLED$th LED through different PDs. In \eqref{intCorr}, while the first term of the summation is the noise correlation matrix, the second term is the interference correlation matrix. The expression $(1-\LEDConnectFlag[\indexUser][\indexLED])$ ensures that only interference signals will be added, because $\LEDConnectFlag[\indexUser][\indexLED]$ is the assignment flag and if $\indexLED$th LED is assigned to $\indexUser$th user, $(1-\LEDConnectFlag[\indexUser][\indexLED])$ is equal to zero. 

\begin{figure}[tb]
	\centering
	\includegraphics[width = 3.45 in]{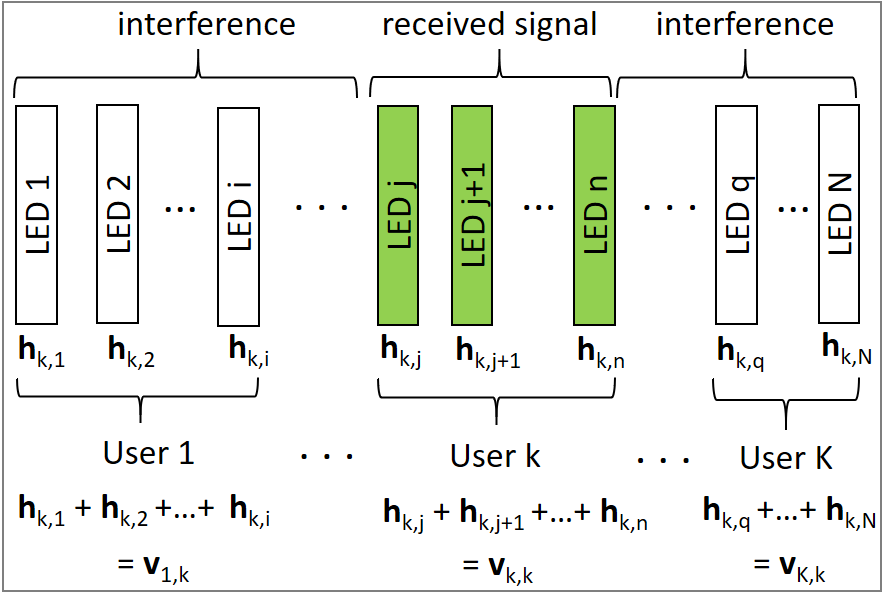}
	\caption{\small Calculation of weights for $\indexUser$th user for OC.}
	\label{OCexplanation}
	\vspace{-3mm}
\end{figure}

\subsection {OC with Known Grouping Information}
If we calculate the interference without considering assignment information, we will ignore the correlation between LEDs that are transmitting to the same interfering user. When the grouping information between interfering LEDs is known, rather than using \eqref{intCorr}, $\textbf{R}$ for the $\indexUser$th user can be calculated as
\begin{equation}
\textbf{R} = N_0B\textbf{I} + \displaystyle\sum_{\substack{\ell = 1 \\ \ell \neq k} }^{\NumberOfUsers}{E[\signalVector[k][\ell]\signalVector[k][\ell]^T]},
\end{equation}
where $\textbf{I}$ is the identity matrix,
\begin{align}
\signalVector[k][\ell] = [\ReceivedPower [\indexUser][1][\ell], \ReceivedPower [\indexUser][2][\ell], ... ,\ReceivedPower [\indexUser][\NumberOfPDs][\ell]]
\end{align}
is the received interference signal from the LEDs which are transmitting to $\ell$th user, and
\begin{align}
\ReceivedPower [\indexUser][\indexPD][\ell] = \PDresp\sum_{\indexLED = 1}^{\NumberOfLEDs}{\LEDConnectFlag[\ell][\indexLED]\power[\indexLED]\channel[\indexUser(\indexPD)][\indexLED]},
\end{align}
is the received interference signal from $\ell$th user at $\indexUser$th user's $\indexPD$th PD. We will refer this method as grouping based OC (GB-OC).

Fig. \ref{OCexplanation} helps to further explain the differences between the calculation of the OC and the GB-OC. Assuming LEDs are numbered with respect to assignment to the users, green LEDs are showing the LEDs assigned to $\indexUser$th user, and white LEDs shows the LEDs assigned to other users which are interference sources for $\indexUser$th user. The vector $\channelVector [\indexUser][\indexLED]$ includes RSS values transmitted from the $\indexLED$th LED and received at different PDs of the $\indexUser$th user. As shown at the bottom of Fig. \ref{OCexplanation}, $\signalVector[\indexUser][j]$ is the sum of $\channelVector [\indexUser][\indexLED]$s transmitted from LEDs that are assigned to $j$th user. While $\signalVector[\indexUser][\indexUser]$ includes the desired signals received at user $\indexUser$, $\signalVector[k][j]$s for $j\neq k$ includes interference. To calculate interference-plus-noise correlation matrix $\textbf{R}$, classical OC sums $\channelVector[\indexUser][i]\channelVector[\indexUser][i]^T$s for interference LEDs, while GB-OC sums  $\signalVector[k][j]\signalVector[k][j]^T$s for $j\neq k$. Classical OC correlation matrix values are lower because cross elements from multiplication of the sum are missing. Therefore, classical OC weight calculation does not include the interference correlation caused by the simultaneous transmission from multiple sources.

\subsection{Relaying Global LED Assignment to Users}
In order for a receiver to implement GB-OC, the LED assignments to all individual users need to be known by each user, which is characterized by the sparse matrix $\textbf{A}$ in~\eqref{Eq:matrixA}. Since all users needs the same information, LED assignment information can be simultaneously broadcast from all the LEDs. The broadcast should be done right after a new LED assignment is computed, and before users start being served with the new assignment.

In the simplest approach, the assignment matrix $\textbf{A}$ can be broadcast from the LEDs to users, which corresponds to an overhead of $N\times K$ bits. Due to the sparse nature of the matrix $\textbf{A}$ (each LED serving only one user), each column of $\textbf{A}$ can be replaced by a single bit vector of size $\lceil\log_2(K+1)\rceil$ representing the identity of the user that is served by a particular LED, and a bit sequence of all zeros if the LED does not serve any user. This second approach therefore requires an overhead on the order of $N\times\log_2K$. LED assignment overhead can be further reduced by grouping LEDs and assigning them to users in groups, which is reminiscent to group-based resource block assignment using the bitmap type-0 and type-1 to reduce control channel overhead in LTE networks~\cite[Table~9.4]{sesia2011lte}.

\begin{figure}[tb]
	\centering
	\includegraphics[width = 3.7 in]{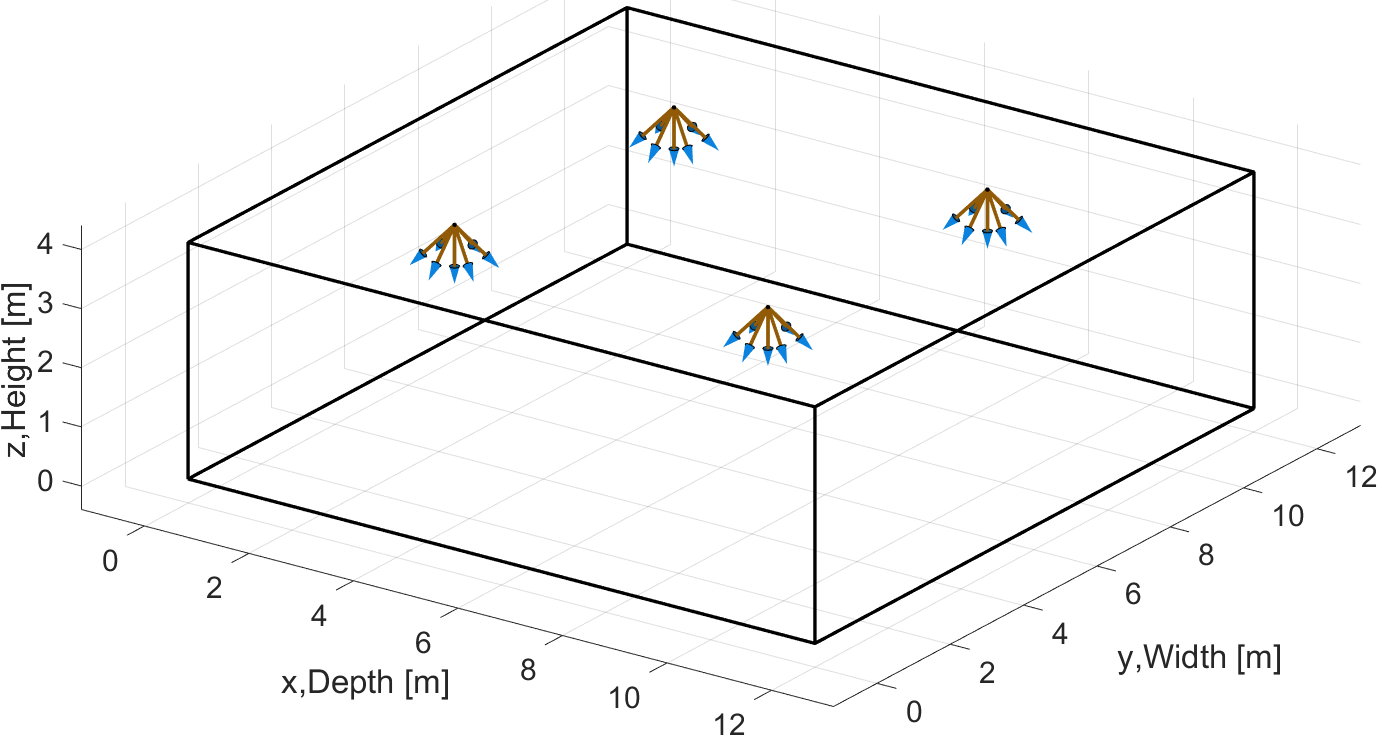}
	\caption{\small The 12m by 12m room and the transmitter locations for simulation evaluations.}
	\label{SimRoom}
\end{figure}

\begin{table}
	\caption{Simulation parameters.}
		\vspace{-2mm}
	\begin{center}
		\begin{tabular}{ |c|c| }
			\hline
			LED directivity index, $\mode$ & 7.0459 \\
			\hline
			{Maximum power coefficient} of an LED, $\power [\rm max]$ & 1 W \\
			\hline
			Responsivity, $\PDresp$ & 0.5 A/W \\
			\hline
			Modulation bandwidth, $B$ &  20 MHz \\
			\hline
			AWGN spectral density, $N_0$ & \small $2.5\times10^{-20}$ A$^2$/Hz \\
			\hline
			Effective surface area (single PD Rx), $A_{\rm R}$ & 40 mm$^2$ \\
			\hline
			Effective surface area (7 PD Rx), $A_{\rm R}$ & 10 mm$^2$ \\
			\hline
			Reflection coefficient (walls)  &  0.8 \\
			\hline
			Reflection coefficient (floor, ceiling)  &  0.3 \\
			\hline
		\end{tabular}
		\label{Table}
	\end{center}
\end{table}

\section{Simulation Results}
For simulations, a square room with dimensions 12~m $\times$ 12~m $\times$ 4~m is considered as in Fig. \ref{SimRoom}. Four multi-element transmitters are considered, each having seven LEDs. In a multi-element transmitter, while one LED is directed downwards, there is a second layer of LEDs having a 45$\degree$ divergence angle with the center LED. The transmitters are located at the ceiling facing downwards; the receivers are assumed to be at 0.85 m height and facing upwards. Other simulation parameters are provided in Table~\ref{Table}. Up to four multipath reflection order, i.e., $\degreeOfReflection = 4$, are considered which are generated using the method discussed in Section II.C. Users are placed at random locations in the room, and the sum rates or fairness indices are calculated over a large number of realizations. For fairness criteria, we present Jain's fairness index (JFI) which takes values between $1/K$ and 1 for $K$ users and the index is given by
\begin{align}
{\rm JFI} = \frac{(\sum_{\indexUser = 1}^{\NumberOfUsers}R_{\indexUser})^2}{\NumberOfUsers\sum_{\indexUser = 1}^{\NumberOfUsers}R_{\indexUser}^2},
\end{align}
where a larger index means a more fair distribution.

\begin{figure}[htp]
	\centering
	\vspace{1mm}
	\subfigure[Sum rate.]{
		\includegraphics[width=3.1in] {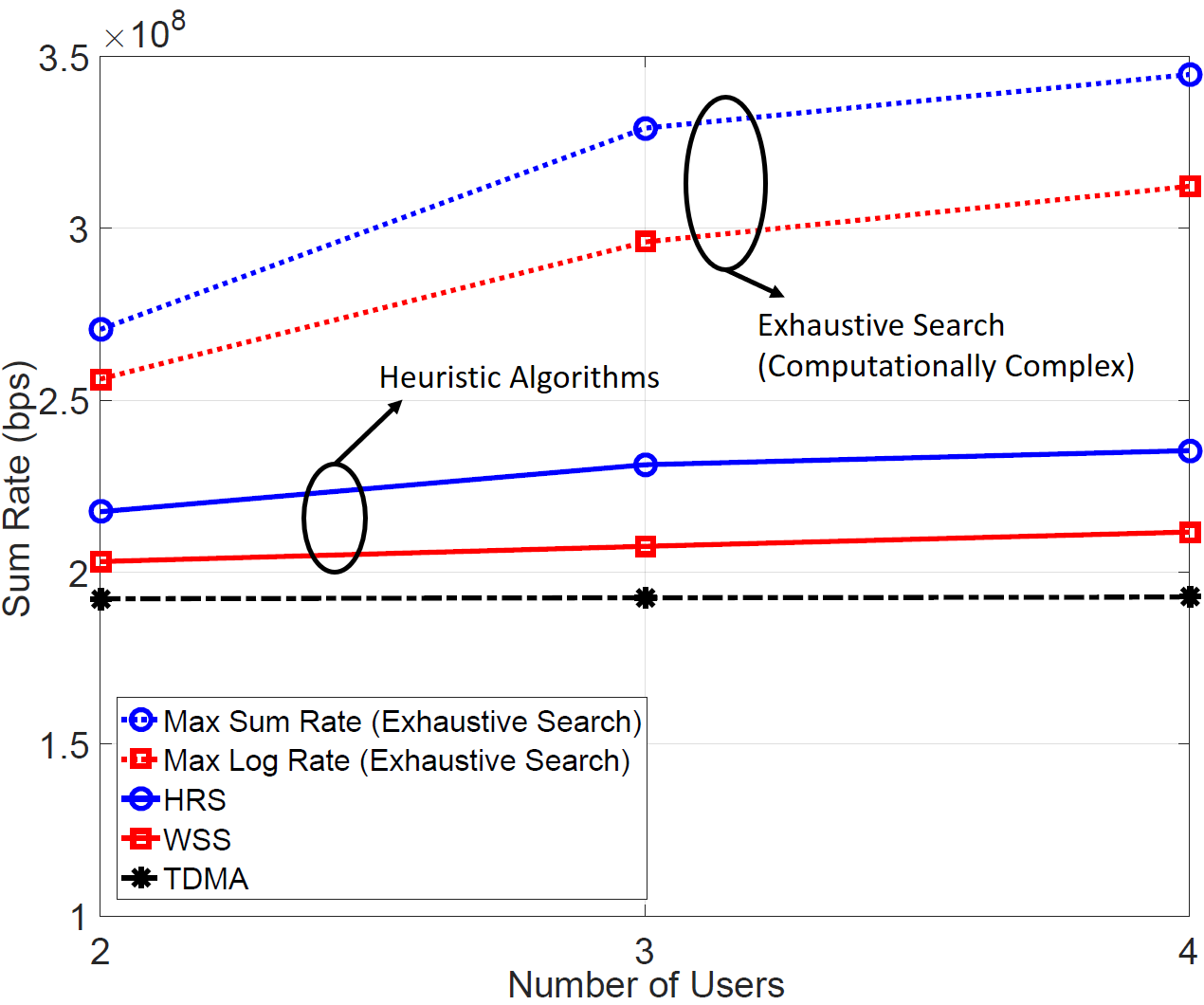}
		\label{RateSmall}
	}
	\subfigure[Logarithmic sum rate.]{
		\includegraphics[width=3.1in] {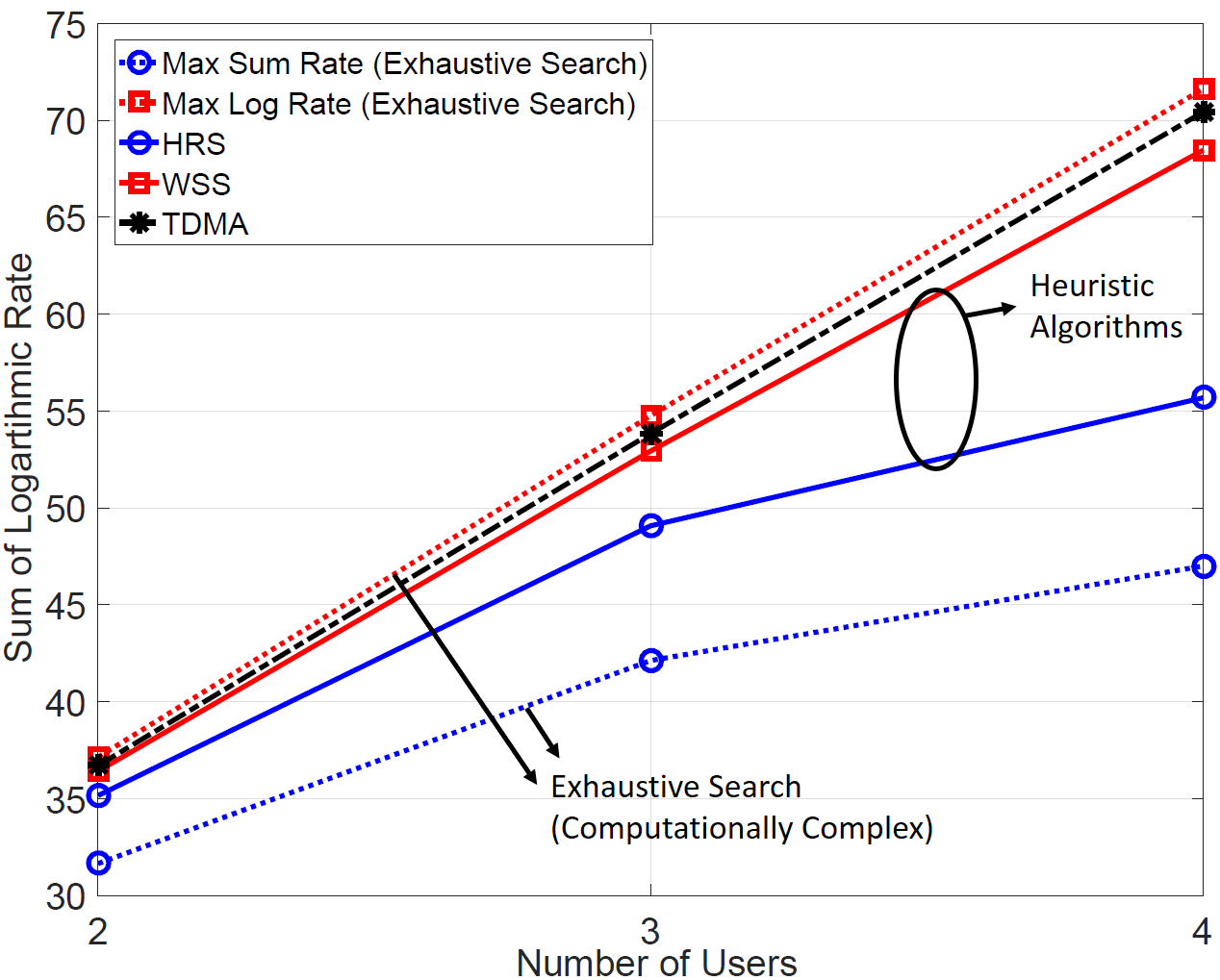}
		\label{LogRateSmall}
	}
	\subfigure[Fairness index.]{
		\includegraphics[width=3.1in] {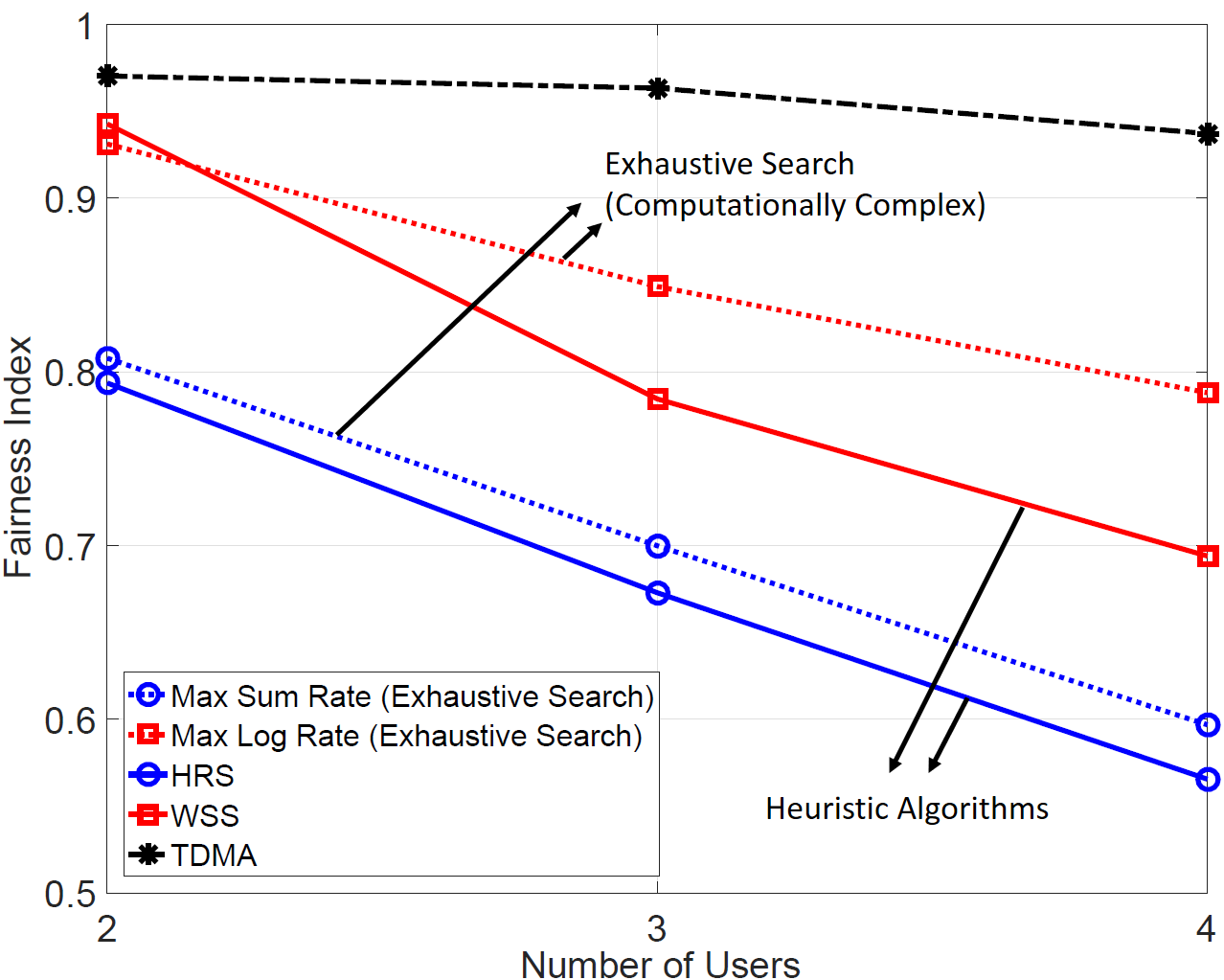}
		\label{FIsmall}
	}
	\caption{Sum rate, sum of logarithmic rate, and JFI for different assignment schemes (including exhaustive search) and different number of users simulated in a small scale scenario (12~m~$\times$~6~m room size with two multi-element transmitters).}
	\label{SmallSetup}
\end{figure}

\subsection{LED Assignment without QoS Constraints (Small Room)}

In Fig.~\ref{SmallSetup}, the sum rate and fairness performance of the proposed assignment schemes and the optimal assignment schemes, which are found by exhaustive search, are shown. For the simulations in Fig.~\ref{SmallSetup}, a room with half the size of the room in Fig.~\ref{SimRoom} is considered with dimensions 12 m $\times$ 6 m. Two multi-element transmitters are used instead of four, and up to four users are simulated. The reason for that is exhaustive search takes exponentially longer time for additional number of LEDs and users, which makes it hard to simulate the scenarios with large number of elements.

In Fig. \ref{SmallSetup}(a), sum rates are given for the \textit{maximum sum rate} assignment that solves \eqref{maxthroughput} by exhaustive search, the \textit{maximum log rate} that solves \eqref{logthroughput} by exhaustive search, the proposed HRS and WSS assignments, and time-division-multiple-access (TDMA). In Fig. \ref{SmallSetup}(b), the sum of log of the rate of the users for the same assignment schemes are given. In Fig. \ref{SmallSetup}(c) the fairness indices are shown. In TDMA case, all LEDs send the same signal and serve one user at a time. All users are served by time division among users with equal length of time slots. 

As expected, maximum sum rate gives the highest rate in Fig. \ref{SmallSetup}(a) and maximum log rate gives highest logarithmic rate in Fig. \ref{SmallSetup}(b). In general, WSS is comparable to maximum logarithmic throughput assignment since they both prioritize proportional fairness criteria, and HRS is comparable to maximum sum rate assignment since they both prioritize maximization of sum rate. HRS and WSS algorithms provide lower sum rate in comparison to the corresponding optimal assignment schemes; however, they both outperform TDMA in terms of data rate. Since the proposed assignment scheme makes use of space diversity, in a larger room or with higher number of users the rate gain over TDMA is expected to be higher (see Fig.~\ref{LargeSetup}). In terms of logarithmic rate, which is a parameter for both higher rate and fair distribution, WSS gives close results to the maximum log rate and TDMA, and it is followed by HRS with a larger margin. Maximum sum rate assignment fails the logarithmic rate criteria and provides the lowest results. In terms of the JFI criteria, WSS provides close results to the max log rate and HRS provides close results to the max sum rate. 

\subsection{LED Assignment without QoS Constraints (Large Room)}

\begin{figure}[htp] 
	\centering
	~\subfigure[Sum rates for different number of users.]{
		\includegraphics[width=3in] {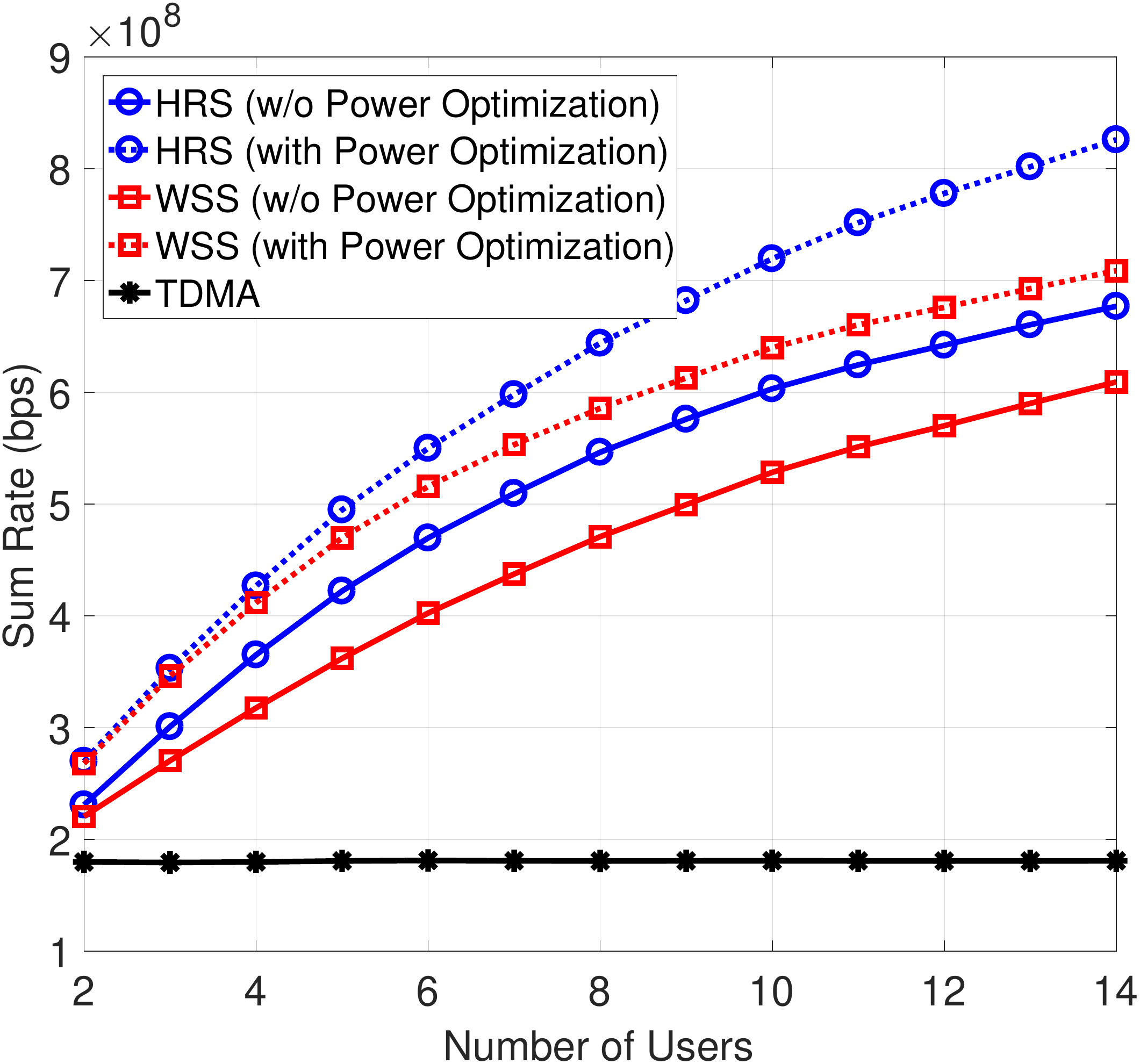}
		\label{Rate}
	}
	\subfigure[Sum of logarithmic rates for different number of users.]{
		\includegraphics[width=3.1in] {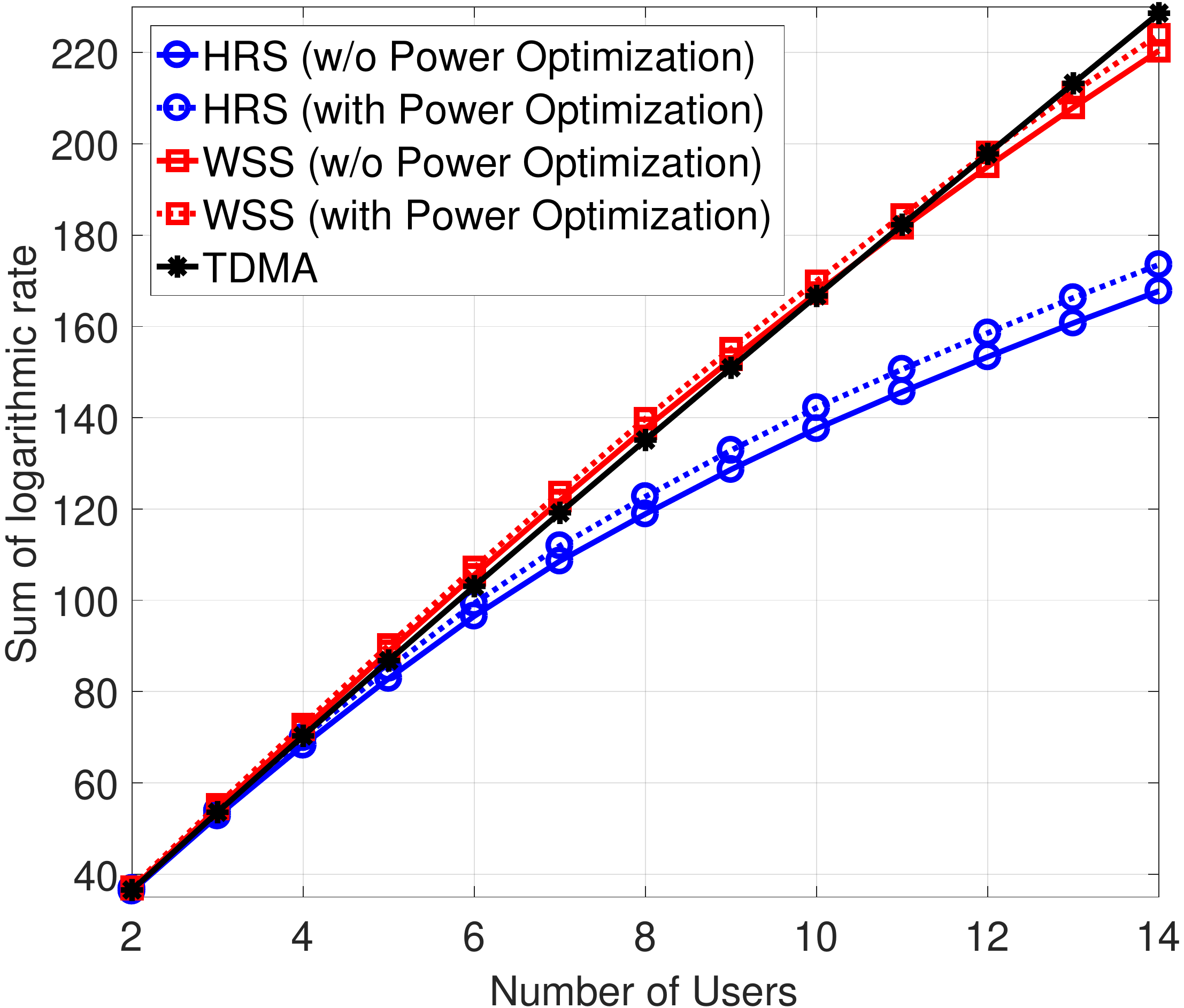}
		\label{LogRate}
	}
	\subfigure[Fairness indexes for different number of users.]{
		\includegraphics[width=3.1in] {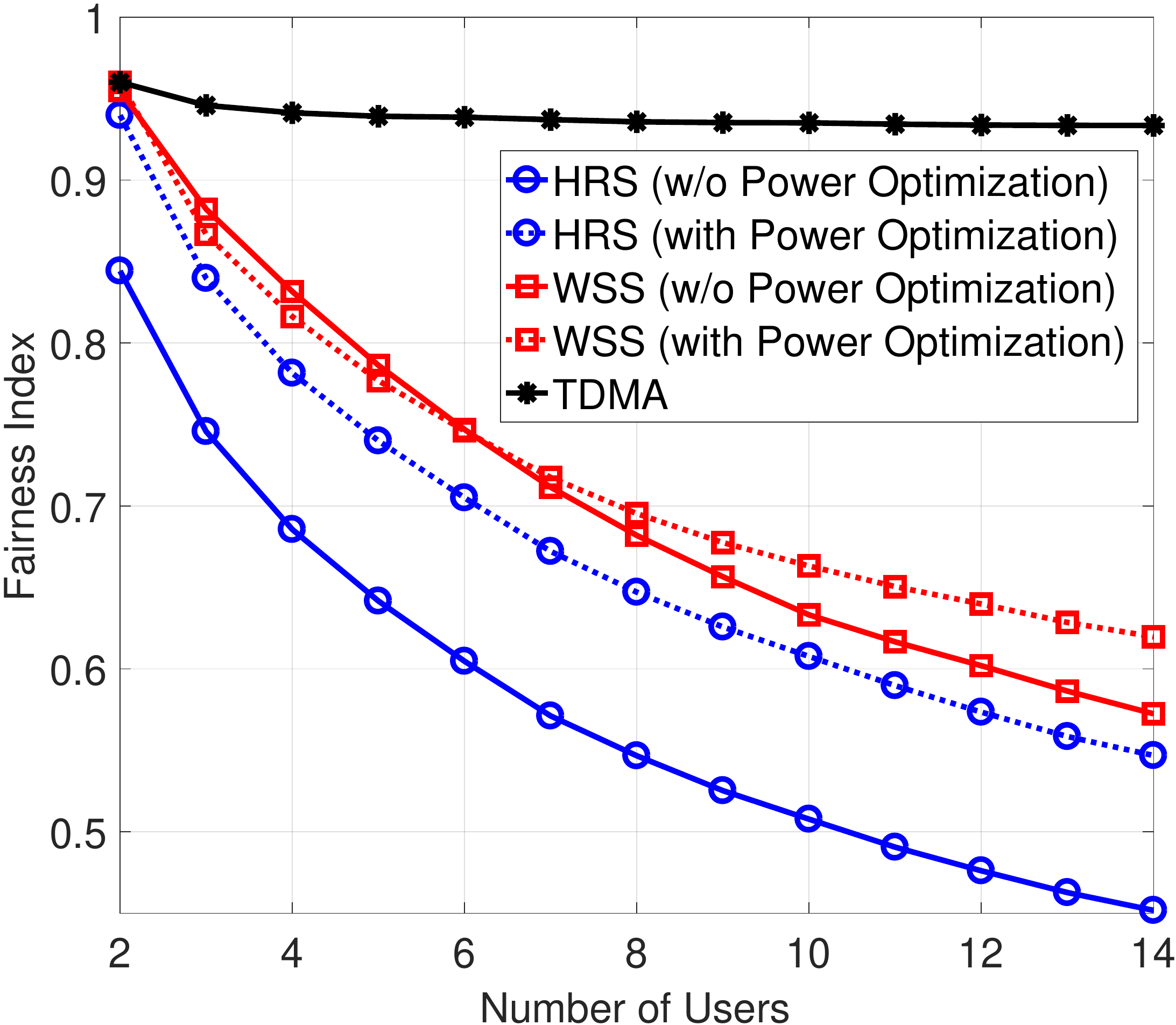}
		\label{FI}
	}
	\caption{  Sum rate, sum of logarithmic rate, and JFI for different assignment schemes and different number of users (12~m~$\times$~12~m room size with four multi-element transmitters as in Fig~\ref{SimRoom}).}
	\label{LargeSetup}
\end{figure}

In Fig.~\ref{LargeSetup}, the room setup in Fig. \ref{SimRoom} is used, where exhaustive simulation results are excluded due to extensively long (on the order of months) simulation duration with today's high performance computers, even with as low as four users. In Fig. \ref{Rate}, the sum rate for HRS and WSS based algorithms are shown. The results with TDMA are also shown for reference. While HRS shows the highest sum rate performance, WSS has slightly lower rate. As HRS assigns each LED to the user with the highest received signal, and does not consider any fairness criterion, it is expected to yield the highest sum throughput. {Sum rates with optimized power coefficients are also shown with dotted lines. All power optimization simulations in this paper maximize the sum of logarithmic throughput as in~\eqref{logthroughput}, using interior-point method. The results show that optimization provides significant gain in the sum rate of both algorithms.} Both algorithms constantly improve the sum rate when the number of users increase, by making use of spatial diversity. When there are 8 users, the proposed algorithms provide more than three times gain over TDMA in terms of data rate. When there are 14 users, the gain reaches approximately five times that of TDMA.  

In Fig. \ref{LogRate} the sum of logarithmic rate performances are shown. WSS shows a similar performance as TDMA by means of logarithmic sum, while HRS yields lower results, especially for higher number of users. {Power optimization slightly increases logarithmic sum rate of both algorithms.} In Fig. \ref{FI}, the fairness index for the same algorithms is given. The WSS provides significantly higher fairness index in comparison to the HRS, since it considers the whole received signal power by a user and provides a more fair LED assignment. {Power optimization increases the fairness index of the HRS significantly, since it maximizes the logarithmic sum rate. Contribution of power optimization to the fairness index of WSS is not that significant. It even causes fairness index to decrease for low number of users. The reason is that, WSS already provides a high fairness index before optimization, and improving logarithmic sum slightly may not improve the fairness index of WSS in all cases. The advantage of power optimization on WSS is mostly visible on the sum rate.}

\begin{figure*}[tp]
	\centering
	\subfigure[Histogram of power coefficients after optimization. The x-axes show the power values, and y-axes show total number of users taking those values ($p_{\rm max}$ = 1 W).]{
		\includegraphics[width=3.1in] {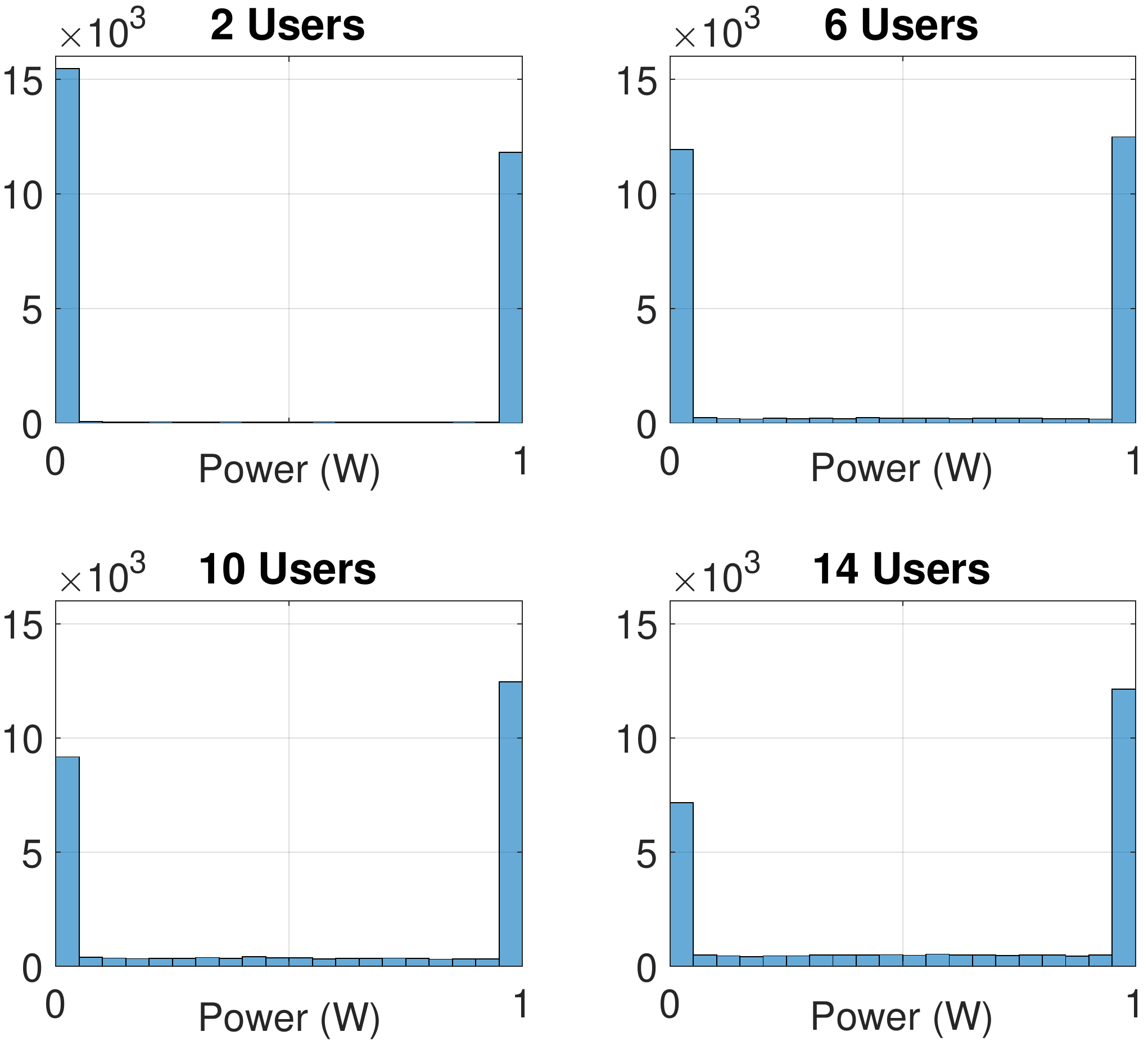}
		\label{Histogram}
	}
	\subfigure[Power coefficients of LEDs for different $p_{\rm max}$ values in a single optimization realization. Different users are shown with different markers.]{
		\includegraphics[width=3.1in] {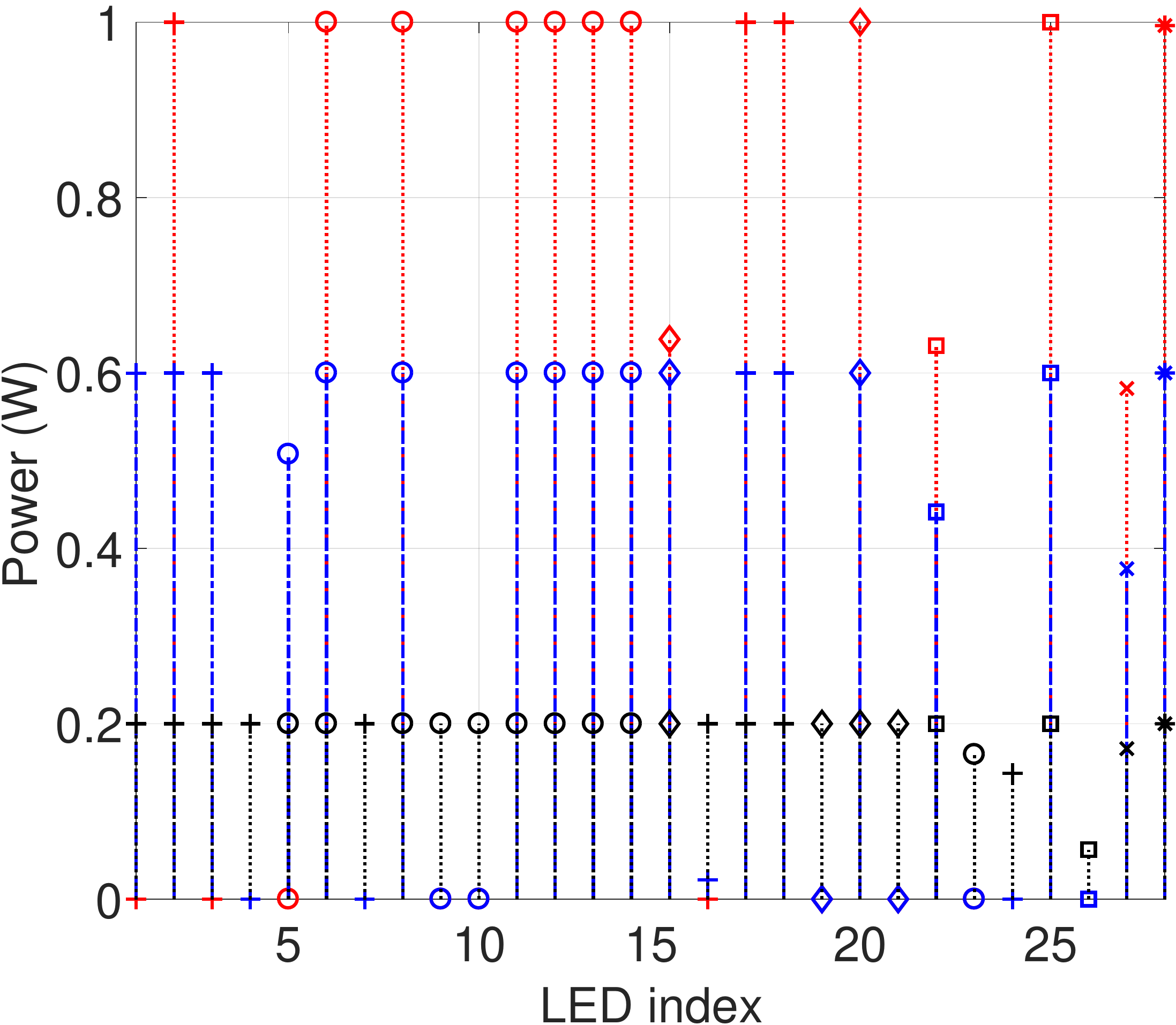}
		\label{PowerLevels}
	}
	\caption{Power coefficients after optimization.}
\end{figure*}

\subsection{Power Control and Illumination}
 
Fig.~\ref{Histogram} shows the histogram of the power coefficients of the simulations in Fig.~\ref{LargeSetup} for WSS (with Power Optimization) case. The histograms shows that the optimum power coefficient tends to be either 1 or 0, depending on if the LED provides more signal power or more interference (to other users), respectively. A similar power control problem is studied in \cite{4600228} under multiple interfering RF links, where they identify scenarios which binary power control is optimal. It might as well be the case in VLC scenarios with high number of LEDs, which we leave as a future study to investigate. In Fig.~\ref{Histogram}, when the number of users increases, lower number of LEDs takes 0 power, which is probably due to decreasing LEDs per user. Since the optimization maximizes sum of logarithmic throughput, it makes sure every user is served by LEDs that are assigned a power coefficient larger than zero. 

In case of dimming, $P_{\rm ave}$ needs to be decreased and it may also limit $p_{\rm max}$. In order to evaluate effects of dimming on power optimization, Fig.~\ref{PowerLevels} shows the power levels of all LEDs for a single optimization realization, for different $p_{\rm max}$ values, that are 1~W, 0.6~W, and 0.2~W. The WSS algorithm is used to assign LEDs to six users, and different users are shown with different markers. While some users are assigned a single LED, some others are assigned as much as 10 LEDs, depending on the distribution of  users in the room. For $p_{\rm max} = 1$~W, many LEDs are assigned a power coefficient of zero, which can also be seen from the histograms in Fig.~\ref{Histogram}. When $p_{\rm max}$ is decreased to 0.2~W, no LED is assigned zero power, and most LEDs are assigned the maximum power value of $p_{\rm max}$. This shows that for lower values of $p_{\rm max}$, noise becomes the dominant factor rather than interference, which makes $p_{\rm max}$ optimal for most LEDs.

\subsection{LED Assignment with QoS Constraints}
In Fig. \ref{PRA}, the sum rate with the PRA algorithm for two different QoS ratios (defined in \eqref{propAssign}) are compared with TDMA. For the results with square marker, half of the users' QoS ratios are 5, and the remaining ones have a ratio~of~1. With TDMA, users can be provided proportional rates by assigning them proportionate time. However, after averaging over large number of realizations, the sum rate is the same for any ratio allocation among users, since users are randomly located at each realization. The PRA provides higher sum rate with respect to TDMA for any number of users and any QoS ratio distribution. As the number of users increases, the capacity gain also increases with respect to TDMA. As mentioned before, this is due to increased spatial diversity between LEDs and users.

\begin{figure}[tb]
	\centering
	\includegraphics[width = 3.4 in]{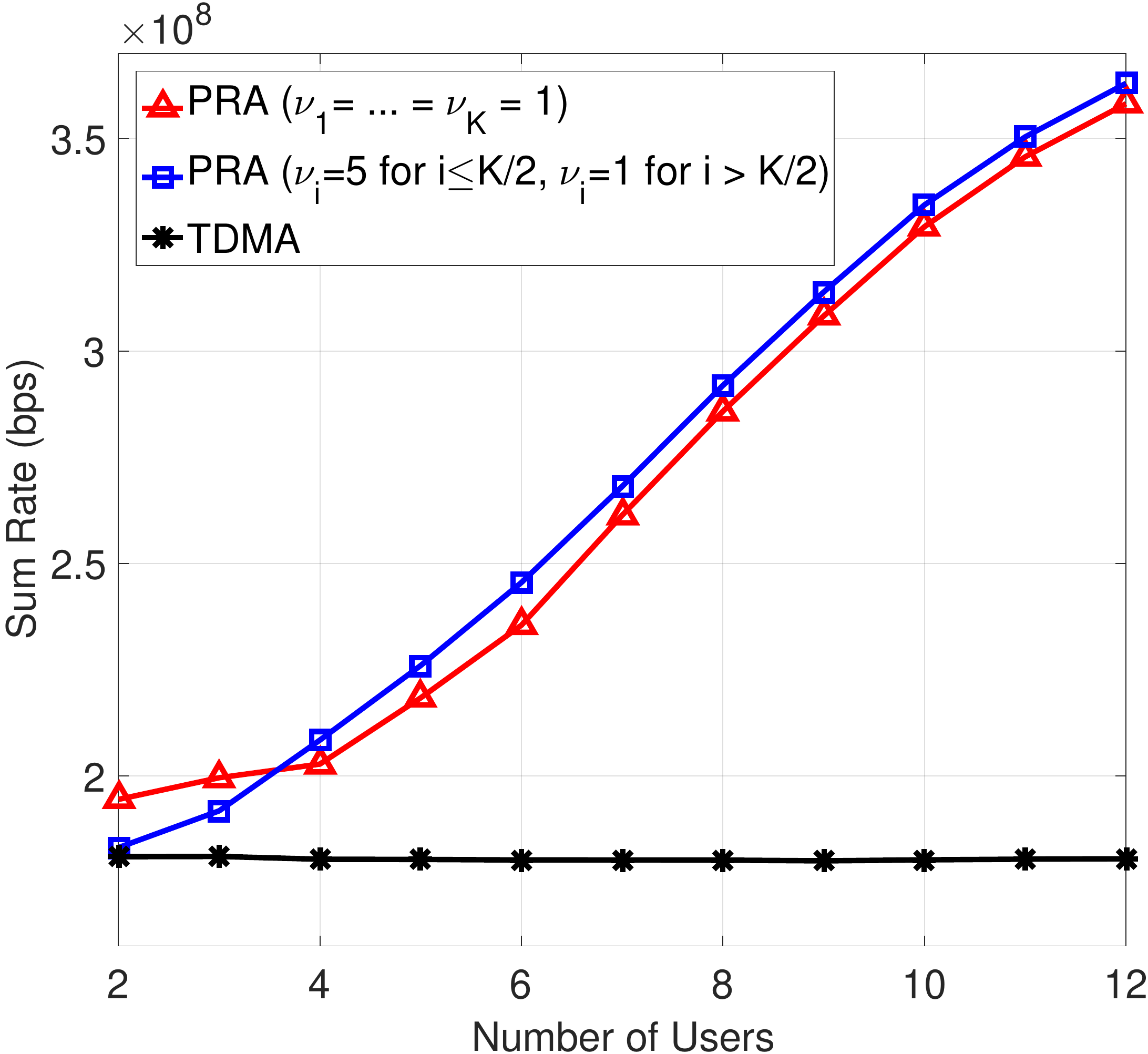}
	\caption{Sum rate for PRA and TDMA for different number of users.}
	\label{PRA}
	\vspace{-2mm}
\end{figure}

With PRA algorithm, QoS ratio difference between users affects sum throughput negatively when the number of users is less than four. In this case, the distribution with equal ratios gives higher sum rate. However, with higher number of users, different QoS ratios do not affect the sum rate negatively. It even provides a slight gain over equal ratio distribution among users. One of the reasons for this behavior is that, since the SINR is proportional to the square of the received signal strength at the receiver, assigning more LEDs on some of the users may increase the rate of those users more than the amount of decrease at the other users.

\subsection{Diversity Combining}
In Fig. \ref{OptComb}, the CDF of SINR for different diversity combining schemes are shown for the deployment scenario in Fig.~\ref{SimRoom}. Four receivers with each having 7 PDs are randomly placed in the room and WSS is used to assign LEDs. CDF data is obtained by averaging over large number of realizations. As expected, OC outperforms MRC: While the gain is around 2~dB for low SINR region, it is more than 10~dB for some high SINR realizations. The OC provides higher gain over MRC for higher SINR regions which causes a stepwise CDF. When the assignment information of LEDs are known, SINR can be further improved by including this information in the calculation of OC weights. The GB-OC, which is OC with known assignment information, outperforms classical version, and a 2 dB to 3 dB gain is observed by using the assignment information. 

\begin{figure*}[htp]
	\centering
	\subfigure[CDF of SINR for four user case.]{
		\includegraphics[width=3.1in] {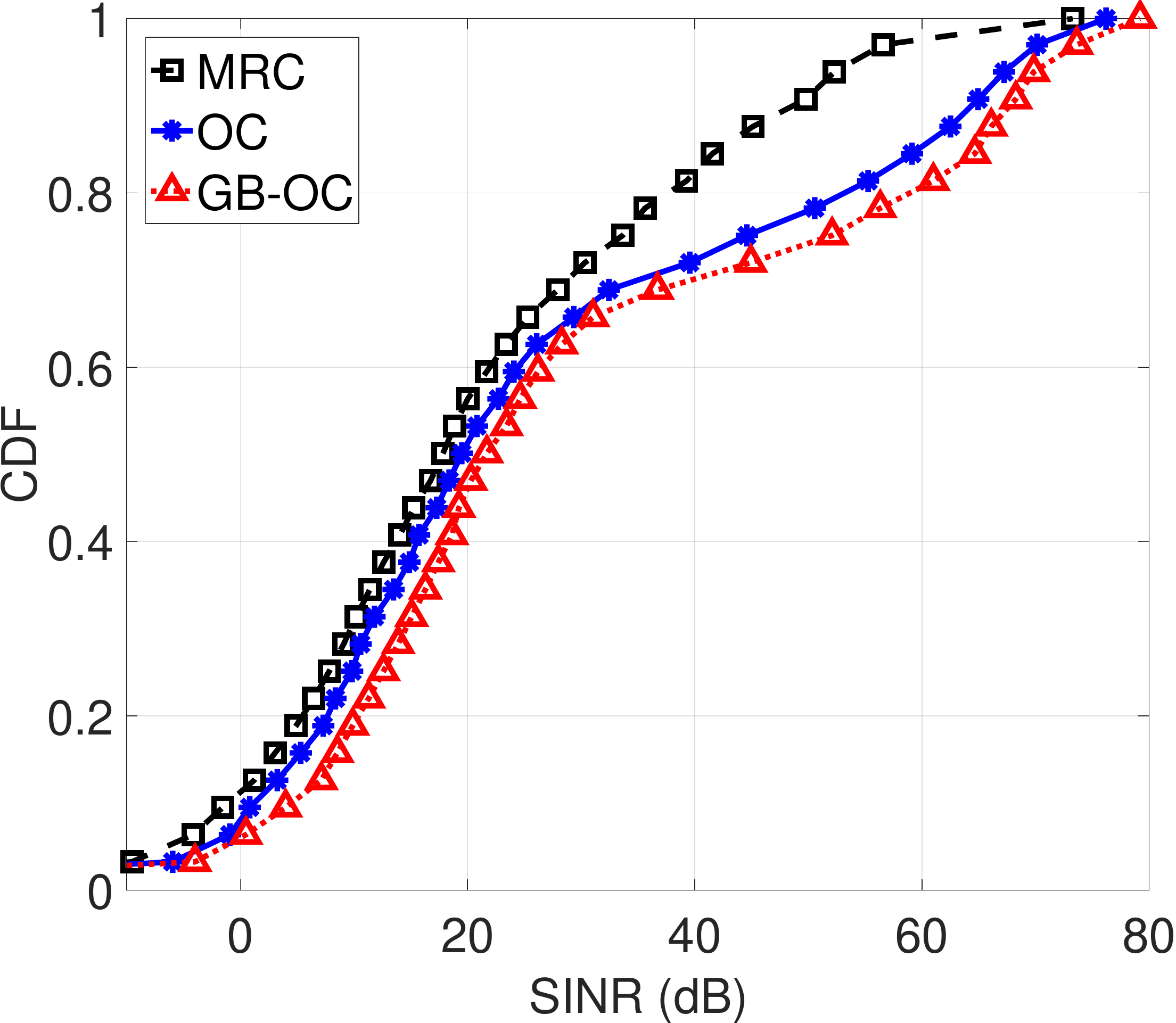}
		\label{OptComb}
	}
	\subfigure[50\% and 10\% SINRs for varying number of users.]{
		\includegraphics[width=3.1in] {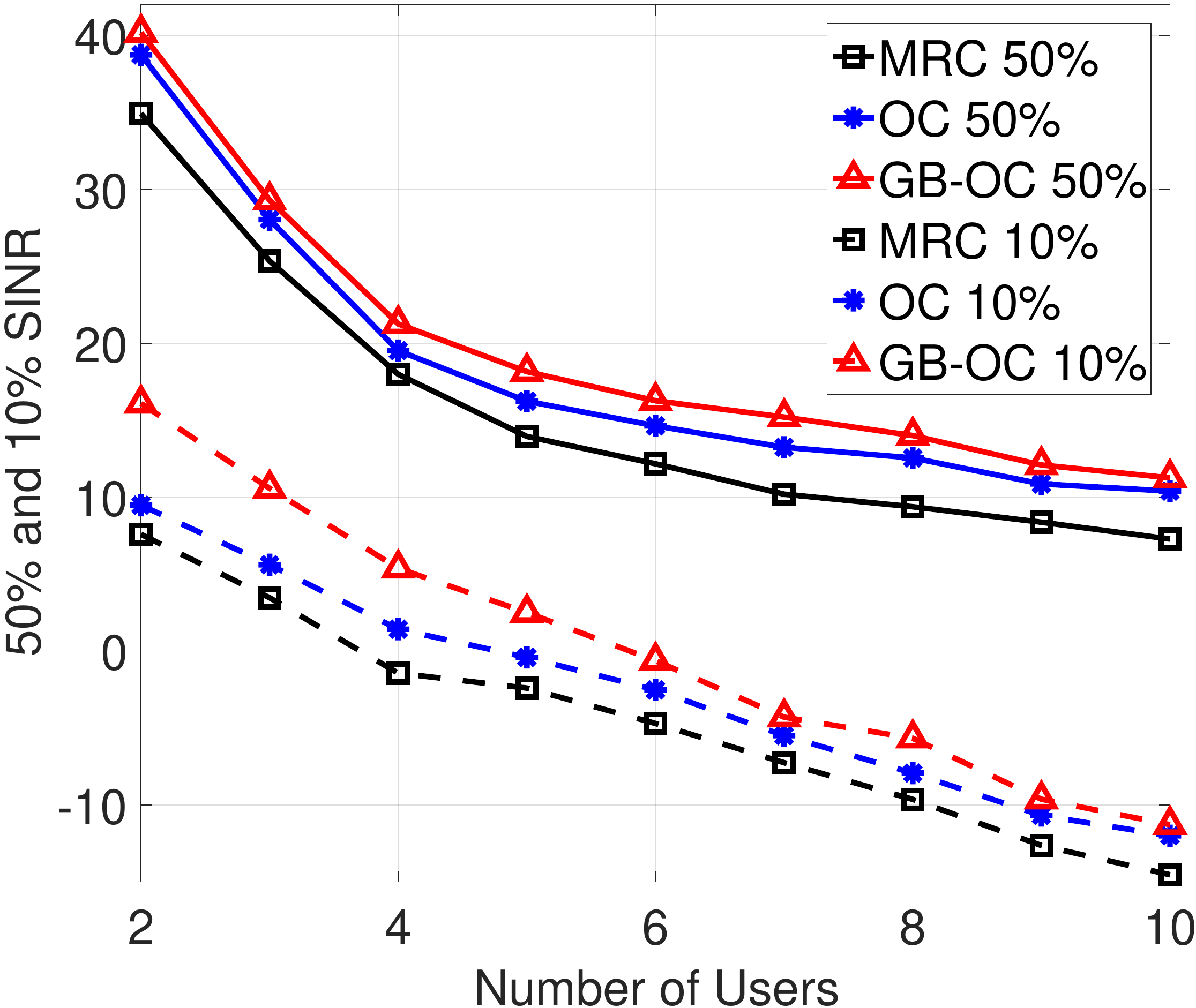}
		\label{OptCombPerc}
	}
	\caption{SINR for different combining techniques.}
\end{figure*}

Fig. \ref{OptCombPerc} shows the 50\% and 10\% of SINR CDF for different number of users, again while utilizing the WSS-based LED assignment. 50\% stands for median SINR and 10\% is for the lowest 10\% of all SINR values. At least a few dB gain is provided by GB-OC for any number of users in both cases. The gain decreases with increasing number of users, especially for 10\% SINR. The reason for that is when there are more users in the room, LED groups are smaller. Therefore coordination caused by grouping of LEDs decreases. We can observe that when the number of users becomes closer to the number of LEDs, GB-OC converges to classical OC. 

In order to get further insights, we plot the average SINR of MRC by location in Fig.~\ref{MRC}, {the SINR difference between OC and MRC at different room locations in Fig.~\ref{OC-MRC}, and the SINR difference between GB-OC and OC in Fig.~\ref{GB-OC-OC}.} To obtain SINR values, similar method as in Fig.~\ref{OptComb} is used. Four users are placed randomly and LEDs are assigned by WSS. The average of 40.000 iterations is considered to decide SINR by location. The total received signal from all LEDs by the receiver is also shown in Fig.~\ref{2Da}. Although the received signal is relatively uniform in the room, the high SINR region of MRC is concentrated under the transmitters. The reason for that is since the user beneath a transmitter is close to many LEDs, it has higher chance to be assigned more LEDs which provides higher signal strength.

\begin{figure*}[htp]
	\centering
	\subfigure[SINR distribution in the room with MRC.]{
		\includegraphics[width=2.6in] {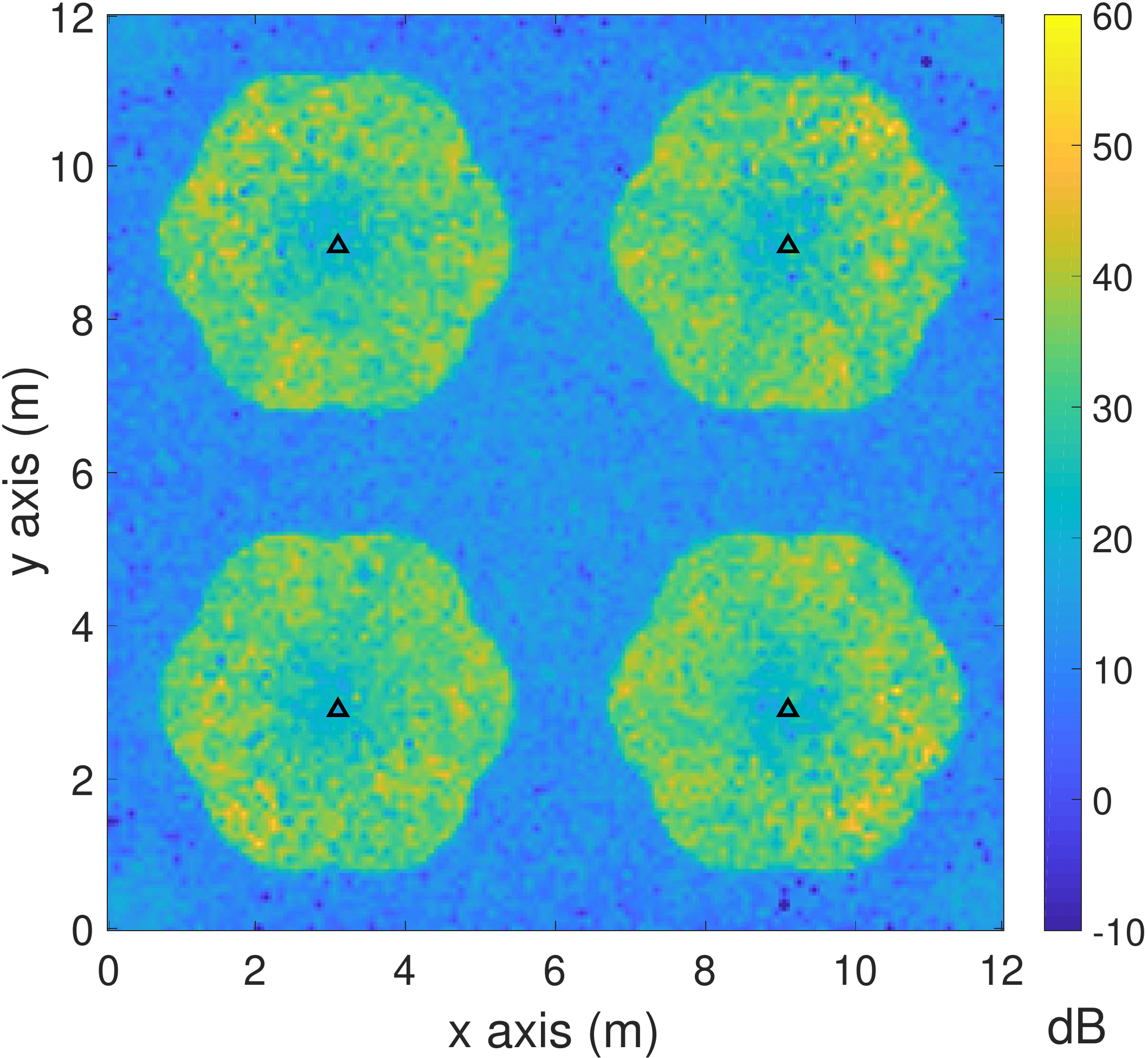}
		\label{MRC}
	}
	\subfigure[  SINR gain of OC over MRC.]{
		\includegraphics[width=2.6in] {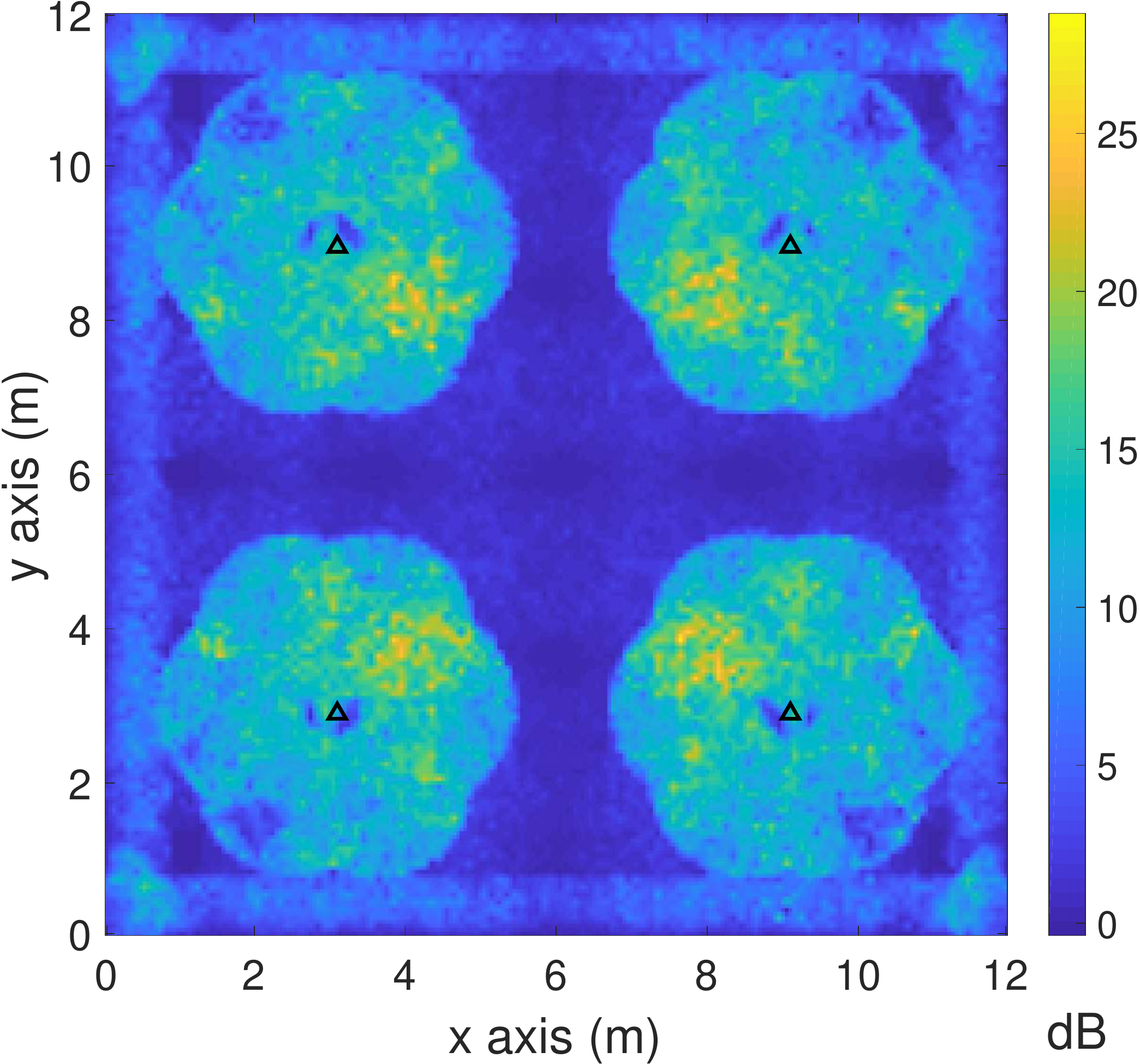}
		\label{OC-MRC}
	}
	\subfigure[  SINR gain of GB-OC over OC.]{
	\includegraphics[width=2.6in] {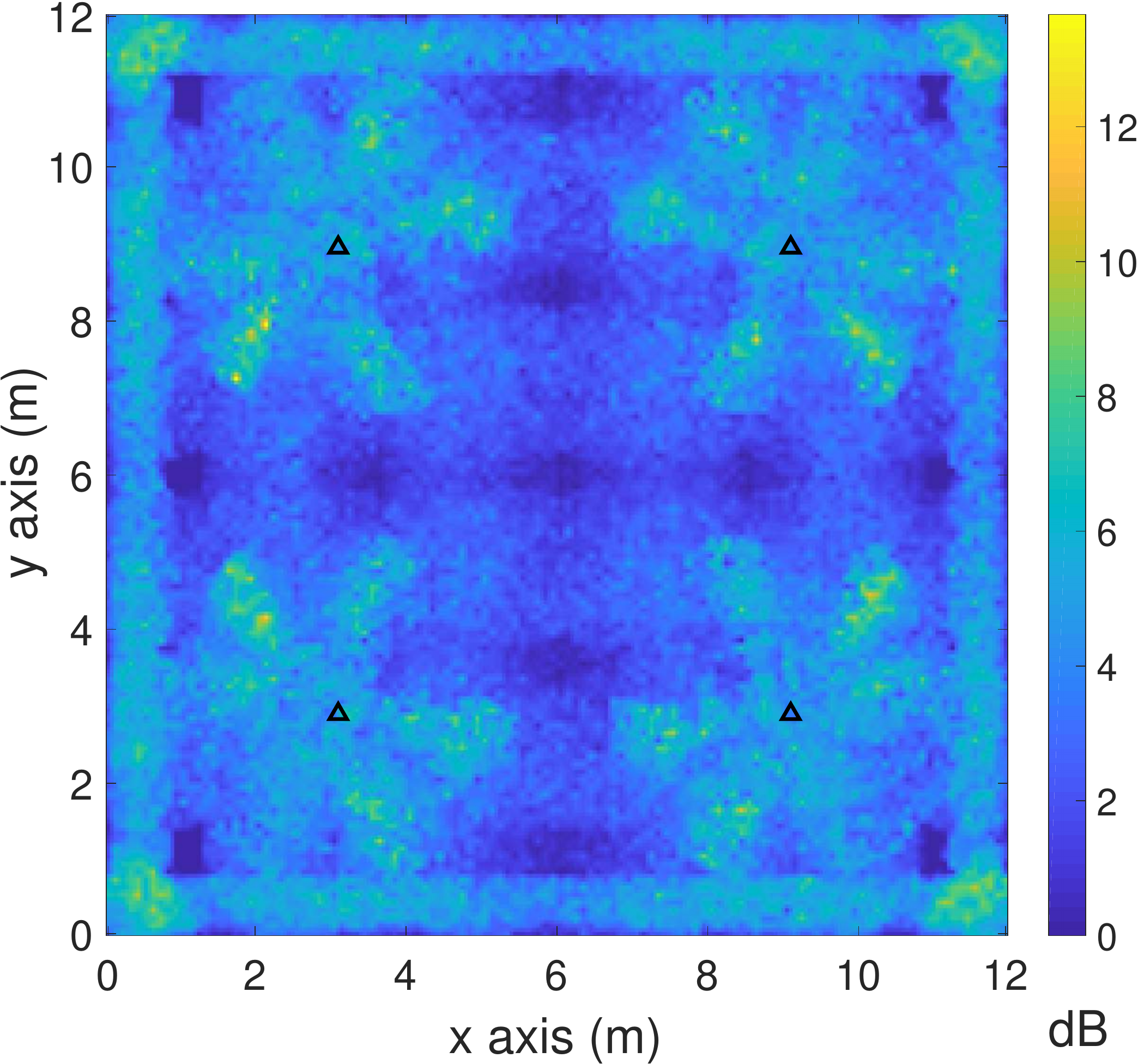}
	\label{GB-OC-OC}
	}
    \subfigure[Total RSS at the receiver (also equal to RSS for TDMA case).]{
	\includegraphics[width=2.6in] {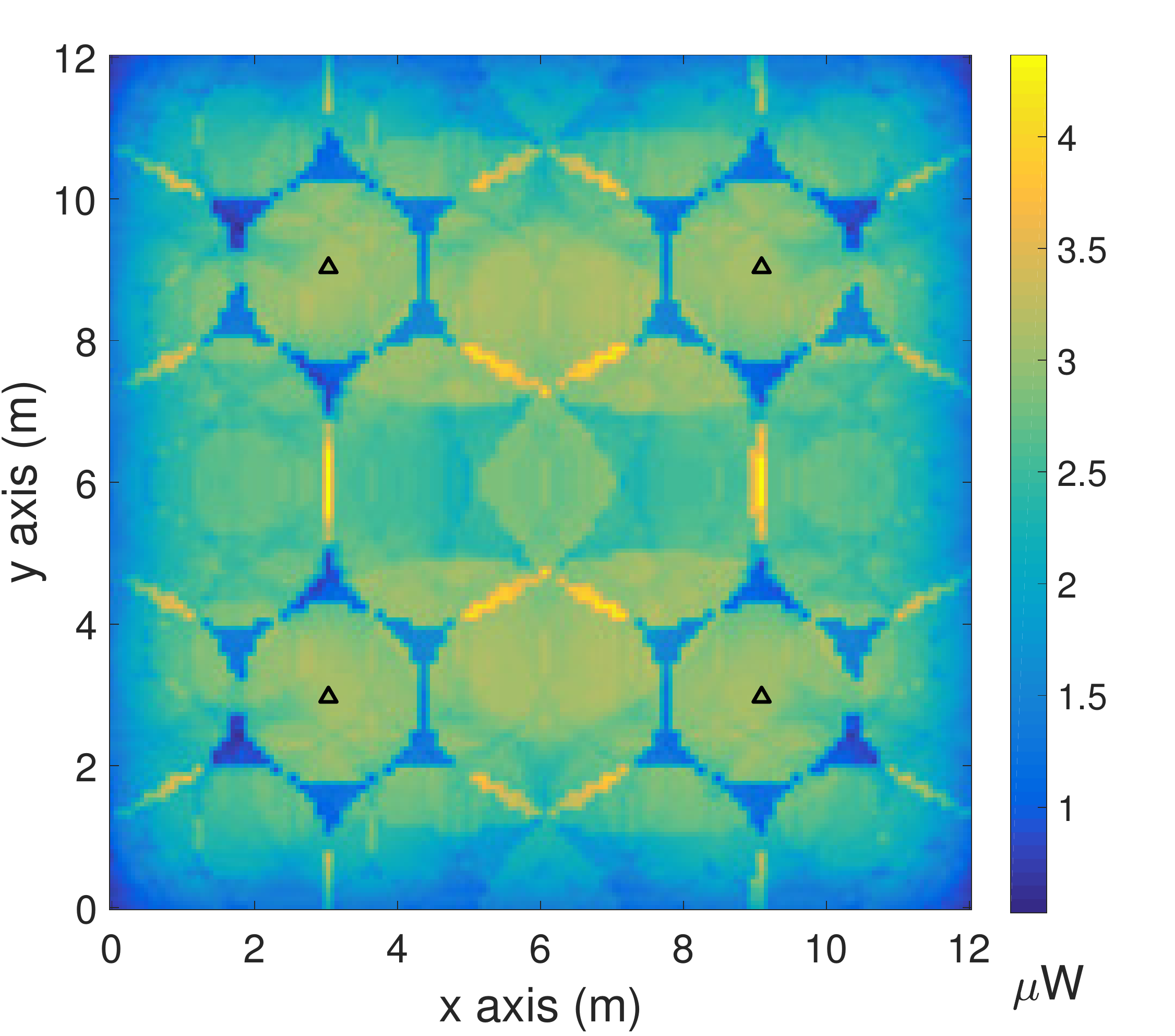}
	\label{2Da}
	}
	\caption{RSS and SINR measurements for different locations within the room of Fig. \ref{SimRoom}. The WSS is used to assign the LEDs to four randomly located users. Triangle markers show the location of the multi-LED VLC access points.}
	\label{2D}
\end{figure*}

Fig.~\ref{OC-MRC} shows that OC provides gain over MRC especially in two different areas. The first one is beneath the transmitters, where users already have high SINR. This behavior results in stepwise CDF by providing additional gain to the high SINR users. In these locations the user is probably assigned all the LEDs of the transmitter above it, and receiving interference signals only from other transmitters, which are at far distance. In this case, since the direction of desired signal and interference is separated, interference correlation is higher which causes OC to perform better. The second area that OC provide higher gain is the location near the walls. This is due to the interference signals caused by wall reflections. The OC suppresses correlated interference reflected from walls and provides higher gain. The gain is doubled at the corners where the reflection signals increase. {Fig.~\ref{GB-OC-OC} shows that while GB-OC provides a relatively uniform gain over OC, the gain follows a similar pattern. That is, the gain is higher beneath the LEDs and near the walls.} 

Fig.~\ref{2Da} shows the total RSS by location, assuming $p_{\rm max}$ power coefficient to all LEDs. This is equivalent to the RSS in TDMA case which all LEDs serve a single user at a time. The total RSS values are also proportional to the illumination level in the room which is perceived by the receiver. Note that the receiver has a FOV constraint and can only receive the light within its FOV. Both the illumination and the RSS have a close to a uniform distribution within the room for the given simulation setup.

\section{Conclusion}
In this paper, we investigate LED-user assignment problem in a downlink VLC scenario, where multiple LEDs serve multiple users. We study suboptimal but computationally efficient LED assignment algorithms {and power optimization} techniques while taking proportional fairness and QoS requirements into account. Our simulation results show that the investigated LED assignment algorithms with equal power distribution among LEDs performs almost as well as the optimal resource allocation schemes. {In addition, power control techniques are shown to provide substantial gains in sum rate and fairness, especially for larger number of users.} We also introduce an improved method for optimal diversity combining at a receiver, taking into account the LED grouping information at the transmitter. With the new approach for calculating the combining weights, SINR gains between 2 dB to 5 dB are obtained in all scenarios. {Our future work includes studying power control with QoS requirements, protocol development/evaluation, and prototyping of the considered multi-element VLC architecture.}  

\section*{Acknowledgment}
The authors would like to thank Bekir S. Ciftler and Wahab Khawaja for reviewing the paper and providing feedback.

\appendices

\section{Derivatives of Rate Function}\label{app:rate_derivatives}
In this appendix, we derive the first and the second-order derivatives of the rate $R_k$. We begin with expressing the SINR expression as $SINR(k)\,{=}\,S_{kk}^2/(T_k{-}S_{kk}^2)$, where the first-order derivatives of $S_{\ell k}$ and $T_k$ with respect to the power coefficients are $\partial s_{\ell k}/\partial p_m\,{=}\,h_{km} \, \delta(\ell,f(m))$ and $\partial T_{k}/\partial p_m\,{=}\,2 s_{\ell k} h_{km} \, \delta(\ell,f(m))$. Taking derivative of $R_k$ in \eqref{Eq:rate} is then given as
\begin{align}\label{eq:rate_1}
\frac{\partial R_{\indexUser}}{\partial p_m} &= \frac{B/\ln 2}{1+SINR(k)}\frac{\partial SINR(k)}{\partial p_m} 
\end{align}
where
\begin{align}
\frac{\partial SINR(k)}{\partial p_m} &= 
\begin{cases}
\displaystyle \frac{2 S_{kk} h_{km}}{T_k-S_{kk}^2}\,, & k{=}\ell\\
\displaystyle {-}\frac{2 S_{\ell k}h_{km}}{T_k-S_{kk}^2} SINR(k)\,,& k{\neq}\ell
\end{cases}\, 
~=~~\frac{2 S_{\ell k}h_{km}}{T_k-S_{kk}^2} C_k^m \,. \label{eq:rate_2}
\end{align}
Realizing $\left[1{+}SINR(k)\right]^{{-}1}{=}\,\left({T_k{-}S_{kk}^2}\right)/T_k$ and employing \eqref{eq:rate_2} in \eqref{eq:rate_1} obtain the first-order derivative of $R_k$ in \eqref{eq:rate_derivative_1st}.\hfill\IEEEQEDhere

The second-order derivative of $R_k$, which is given as 
\begin{align}\label{eq:rate_3}
\frac{\partial^2 R_\indexUser}{\partial p_m \partial p_n} &= \frac{B}{\ln 2} 2 h_{km} \frac{\partial}{\partial p_n} \left( \frac{S_{\ell k}}{T_k} C_k^m \right),
\end{align}
can be evaluated by examining the following two conditions where we employ the derivative of the SINR given in \eqref{eq:rate_2} when necessary.
\begin{case}\label{case:case_equal_l}
Assuming $\ell{=}\ell'$, where $\ell{=}f(m)$ and $\ell'{=}f(n)$, we have
\begin{align}\label{eq:rate_4}
\frac{\partial}{\partial p_n} \left( \frac{S_{\ell k}}{T_k} C_k^m \right) &=
\begin{cases}
\displaystyle h_{kn}\frac{T_k {-} 2S_{\ell k}^2}{T_k^2} \,, & k{=}\ell \\
\displaystyle h_{kn}\frac{SINR(k)}{T_k}  \left( \frac{2S_{\ell k}^2}{ T_k {-} S_{kk}^2} {-}\frac{T_k{-}2S_{\ell k}^2}{T_k} \right)\,, & k{\neq}\ell
\end{cases}
\end{align}

\end{case}

\begin{case}\label{case:case_notequal_l}
When $\ell \neq \ell'$, the desired derivative becomes 
\begin{align}\label{eq:rate_5}
\frac{\partial}{\partial p_n} & \left( \frac{S_{\ell k}}{T_k} C_k^m \right) = S_{\ell k} \frac{\partial}{\partial p_n} \left( \frac{C_k^m}{T_k} \right) =
\begin{cases}
\displaystyle {-}\frac{2S_{\ell k}S_{\ell' k}h_{kn}}{T_k^2} \,, & k{=}\ell \text{ or } k{=}\ell' \\
\displaystyle \frac{2S_{\ell k}S_{\ell' k}h_{kn}}{T_k^2} SINR(k) \Big( 2{+}SINR(k) \Big) \,, & k{\neq}\ell, k{\neq}\ell'  
\end{cases}
\end{align}
\end{case}
\noindent Combining \eqref{eq:rate_4} and \eqref{eq:rate_5} in \eqref{eq:rate_3} obtains the second-order derivative of $R_k$ given in \eqref{eq:rate_derivative_2nd}. 

\hfill\IEEEQEDhere  

\bibliographystyle{IEEEtran} 
\bibliography{JSAC}

\end{document}